\documentclass[a4paper,12pt]{article}

\usepackage[T1]{fontenc}       
\usepackage{amsmath}
\usepackage{graphicx}

\newcommand{\fett}[1]{\mbox{\boldmath$#1$}}
\usepackage{overcite}
\newcommand{\onlinecite}[1]{\hspace{-1 ex} \nocite{#1}\citenum{#1}}

\def\mathcolor#1#{\@mathcolor{#1}}
\def\@mathcolor#1#2#3{%
  \protect\leavevmode
  \begingroup
    \color#1{#2}#3%
  \endgroup
}
\usepackage{braket}

\title{\vspace*{-1cm}\bf Comprehensive Analysis of the Neglect of Diatomic Differential Overlap Approximation}
\author{Tamara Husch and Markus Reiher\thanks{corresponding author: markus.reiher@phys.chem.ethz.ch.}
\vspace{10 mm}\\
ETH Z\"urich, Laboratorium f\"ur Physikalische Chemie,\\ Vladimir-Prelog-Weg 2, 8093 Z\"urich, Switzerland.
}

\begin{document}

\maketitle

\begin{center}
 \textbf{Abstract}
\end{center}
Many modern semiempirical molecular orbital models 
are built on the 
neglect of diatomic differential overlap (NDDO) approximation. 
An in-depth understanding of this approximation is therefore indispensable 
 to rationalize the success of these semiempirical molecular orbital models 
 and to develop further improvements on them.
The NDDO approximation provides a recipe to approximate  
electron-electron repulsion integrals (ERIs) in a symmetrically orthogonalized basis based on a far
smaller number of  ERIs in a locally orthogonalized basis.
We first analyze the NDDO approximation by comparing 
ERIs in both bases for a selection of molecules and for a selection of basis sets.
We find that the errors in Hartree--Fock and  
 second-order M{\o}ller--Plesset perturbation theory  
energies grow roughly linearly with the number of basis functions.
We then examine different approaches to correct for the 
 errors caused by the NDDO approximation and propose a strategy to directly 
correct for them in the two-electron matrices that enter the Fock operator.

\newpage
\section{Introduction}

The neglect of diatomic differential overlap (NDDO) approximation\cite{Pople1965} is 
the foundation of the MNDO model\cite{Dewar1977} and is therefore passed on  
to modern semiempirical molecular orbital (SEMO) models
such as AM1\cite{Dewar1985}, PM$x$ ($x=3,6,7$)\cite{Stewart1989,Stewart2007,Stewart2012}, 
and OM$x$ ($x=1,2,3$).\cite{Kolb1993a,Weber2000a,Dral2016b}
These SEMO models are chosen when, on the one hand, accurate \textit{ab initio} electronic structure models 
are computationally unfeasible ---  but when, on the other hand, a calculation with an 
electronic structure model is favored over a classical force field 
to exploit the first-principles nature of the fundamental electrostatic interactions.
Examples include the simulation of very large systems such as proteins, \cite{Senn2006,Alexandrova2008a,
Senn2009a,Alexandrova2009,Stewart2009,Acevedo2010, Doron2011,Polyak2012} 
 virtual high-throughput screening schemes for materials and drug 
discovery, \cite{Husch2015a, Husch2015,Lepsik2013,S.Brahmkshatriya2013, 
Yilmazer2015, Vorlova2015a,DuyguYilmazer2016,Sulimov2017}
or real-time interactive quantum chemical calculations. \cite{bosson2012,Haag2013,Haag2014,
Haag2014a,Vaucher2016a,Muhlbach2016,Heuer2018} 

Originally, the NDDO approximation was conceptualized as a means to 
 reduce the number of electron-electron repulsion 
integrals (ERIs) that need to be explicitly calculated in the course of a 
Hartree--Fock (HF) calculation. \cite{Pople1965}
The NDDO approximation specifies how ERIs 
in a symmetrically orthogonalized basis may be approximated based on a far
smaller number of  ERIs in a locally orthogonalized basis.
However, the NDDO approximation has not found acceptance 
in \textit{ab initio} calculations due to the significant errors that it 
introduced in ERIs in the symmetrically orthogonalized basis. \cite{Cook1967,Roby1969,Sustmann1969,Gray1970,
Roby1971, Brown1971,Roby1972,Brown1973,Birner1974,Chandler1980,Duke1981, Weinhold1988,Koch2014,
Neymeyr1995a,Neymeyr1995c,Neymeyr1995d,
Neymeyr1995,
Neymeyr1995b,Tu2003}
Instead, the NDDO approximation has become popular in SEMO models where it 
is combined\cite{Dewar1977,Dewar1985,Stewart1989,Stewart2007,Stewart2012,
Kolb1993a,Weber2000a,Dral2016b} with various other approximations made to the ERIs, to the 
one-electron matrix, and to the nucleus-nucleus repulsion energy which 
benefit from mutual error compensation. 
We refer to Ref.~\onlinecite{Husch2018b} for a comprehensive summary of all approximations incorporated in 
popular NDDO-SEMO models. 
An important approximation to highlight in this context is the 
empirical modification of the ERIs in NDDO-SEMO models\cite{Dewar1977,Dewar1985,Stewart1989,Stewart2007,Stewart2012,
Kolb1993a,Weber2000a,Dral2016b} where the ERIs are scaled  
so that they are usually significantly smaller than the 
analytically calculated ones. This scaling then compensates other approximations in SEMO models. 
The results obtained with \textit{ab initio} electronic structure models invoking the NDDO approximation 
 can therefore not be directly related to results obtained with modern NDDO-SEMO models.
Nevertheless, an in-depth analysis of 
the NDDO approximation is mandatory to develop further improvements to NDDO-SEMO models.
Despite decades of work on NDDO-SEMO models,
a fully satisfactory analysis has not been provided yet. 
We intend to take a step toward closing this gap here 
and study the foundations of the NDDO approximation from a state-of-the-art perspective.

First, we determine how the NDDO approximation affects 
 ERIs evaluated in a symmetrically orthogonalized basis. 
Previous analyses of the NDDO approximation were limited to few tens of 
molecules that consisted of a
few atoms (usually less than four heavy atoms).
\cite{Roby1969,Roby1971, Roby1972,Brown1971,Brown1973,Weinhold1988,Koch2014,
Sustmann1969, Birner1974,Duke1981,Tu2003,
Gray1970,Chandler1980,Neymeyr1995,Neymeyr1995a,Neymeyr1995b,Neymeyr1995c,Neymeyr1995d}
In this work, we consider a diverse selection of molecules that are also 
much larger. 
As the errors in the ERIs will propagate to 
all quantities calculated in an NDDO framework, we study 
 how the NDDO approximation affects the (absolute and relative) HF and 
second-order M{\o}ller--Plesset perturbation (MP2) theory  
energies. 
In this context, we examine different basis set choices.

In general, the NDDO approximation is only valid for  
a locally orthogonal basis set, \cite{Neymeyr1995a,Neymeyr1995b,Neymeyr1995c,Neymeyr1995d}
which appears 
to restrict contemporary NDDO-SEMO models to a minimal basis set.
A minimal basis set, however, is generally unsuitable for the description of atoms in
molecules because it does not yield reliable relative energies, force constants, 
electric dipole moments, static dipole polarizabilities, and other properties. \cite{Giese2005,Szabo2012,Helgaker2014,
Kolos1979,Francl1982,Davidson1986} 
It was suggested to generalize NDDO-SEMO models to larger, e.g., double-zeta split-valence 
basis sets to obtain more accurate results. \cite{Dewar1992a}
Two studies examined \cite{Thiel1981a,Gleghorn1982} the effects of the application of 
a double-zeta split-valence basis set in conjunction with the NDDO approximation, 
but came to the conclusion that, contrary to what one would expect, the results did not improve compared to 
the results obtained with a single-zeta basis set.
In this work, we dissect in detail the origins of this counterintuitive observation.

It does not come as a surprise that we find --- in agreement with previous results 
\cite{Roby1969,Roby1971, Roby1972,Brown1971,Brown1973,Weinhold1988,Koch2014,
Sustmann1969, Birner1974,Duke1981,Tu2003,
Gray1970,Chandler1980,Neymeyr1995,Neymeyr1995a,Neymeyr1995b,Neymeyr1995c,Neymeyr1995d} --- 
  the NDDO approximation to cause severe errors.
We therefore examine how one can correct for these errors.
We briefly review the error compensation strategy that 
contemporary NDDO-SEMO models apply and then propose a way to directly 
correct for errors in the two-electron matrices.
We show that our approach allows for rapid calculations invoking the NDDO approximation with error control.

\section{Neglect of Diatomic Differential Overlap Approximation}

\subsection{Basic Notation}

In a basis-set representation, each spatial molecular orbital $\psi_i=\psi_i(\fett{r})$ is approximated 
as a linear combination of pre-defined basis functions. 
Following the well-established approach for finite systems in molecular physics, 
we choose the basis functions to be Gaussian-type atom-centered functions 
$\chi_\mu^I=\chi_\mu^I(\fett{r})$. 
Our notation indicates that the $\mu$-th basis function of type $\chi$ is centered on atom $I$. 
Additionally, we require the basis functions $\chi$ 
to be locally orthogonal which means that 
 the overlap ${}^{\chi}\!{S}_{\mu\nu}$  between the basis functions 
$\chi_\mu^I$ and $\chi_\nu^J$ must fulfill, 
\begin{equation}
\label{eq:locorth}
 \;{}^{\chi}\!{S}_{\mu\nu} = 
\begin{cases}
 \left<\chi_{\mu}^I\middle|\chi_{\nu}^J\right> & I\ne J,\ \forall \mu,\nu \\
 \delta_{\mu\nu} & I=J,\ \forall \mu,\nu \\
\end{cases},
\end{equation}
where $\delta_{\mu\nu}$ is the Kronecker delta.
A molecular orbital $\psi_i$ is then given as the sum of 
the $M$ basis functions $\chi_\mu^I$
weighted with expansion coefficients ${}^\chi\fett{C} =\{ {}^\chi C_{\mu i}\}$,
\begin{equation}
 \psi_i(\fett{r}) = \sum_{\mu=1}^M {}^\chi C_{\mu i} \chi^I_\mu(\fett{r}).
\end{equation}
Throughout this work, we denote the bases by a left superscript, 
i.e., `${}^\chi \fett{C}$'.
In the $\chi$-basis, the Roothaan--Hall equation in the  
spin-restricted formulation then reads
\cite{Szabo2012}
\begin{equation}
 \label{eq:roothanhall_tau}
 {}^\chi {\fett{F}}({}^\chi {\fett{C}})\; {}^\chi {\fett{C}} = 
{}^\chi {\fett{S}} \; {}^\chi {\fett{C}} \fett{\epsilon},
\end{equation}
where ${}^\chi {\fett{F}}$ is the Fock matrix, which depends 
on ${}^\chi\fett{C}$, ${}^\chi {\fett{S}}$ is 
the overlap matrix, and $\fett{\epsilon}$ is the diagonal 
matrix of orbital energies.
As $\fett{\epsilon}$
is invariant under unitary matrix transformations with which we may transform one basis into 
another one, it does not carry a left superscript.
The evaluation of the Fock-matrix elements in the $\chi$-basis, \cite{Szabo2012}
\begin{equation}
\label{eq:fockmatrix_chi}
\begin{split}
 \;{}^{\chi}\!{F}_{\mu\nu}({}^\chi {\fett{C}}) &= \left<\chi_\mu^I|h|\chi_\nu^J\right>
+ \sum_{\lambda\sigma}^M \;{}^{\chi}\!{P}_{\lambda\sigma}({}^\chi {\fett{C}})
\left[ \left<\chi_\mu^I\chi_\nu^J|\chi_\lambda^K\chi_\sigma^L\right> 
                 -\frac{1}{2} \left<\chi_\mu^I\chi_\sigma^L|\chi_\lambda^K\chi_\nu^J\right> \right], \\
\end{split}
\end{equation}
requires the evaluation of one-electron 
integrals $\left<\chi_\mu^I|h|\chi_\nu^J\right>$,
 elements of the density matrix  
${}^{\chi}\!{\fett{P}}({}^\chi {\fett{C}})$, and ERIs in the $\chi$-basis ($^{\chi}$ERIs).
In a spin-restricted framework, 
the elements of ${}^{\chi}\!{\fett{P}}({}^\chi {\fett{C}})$ are given by 
\begin{equation}
\begin{split}
 \;{}^{\chi}\!{P}_{\mu\nu}({}^\chi {\fett{C}}) &= 2 \sum_{i=1}^{n/2}\;{}^{\chi}\!{C}_{\mu i} \;{}^{\chi}\!{C}_{\nu i}, \\
\end{split}
\end{equation}
where $n$ is the number of electrons, 
and the $^{\chi}$ERIs are calculated according to 
\begin{equation}
   \left<\chi_\mu^I\chi_\nu^J|\chi_\lambda^K\chi_\sigma^L\right> =
\int \int 
\chi_\mu^{*,I}(\fett{r}_1) 
\chi_\nu^{J}(\fett{r}_1)
\frac{1}{|\fett{r}_1-\fett{r}_2|}
\chi_\lambda^{*,K}(\fett{r}_2)
\chi_\sigma^{L}(\fett{r}_2) d^3r_1 d^3r_2.
\end{equation}
For the following discussion, it is convenient to divide the Fock matrix into 
a one-electron matrix $^\chi\fett{H}$ and a two-electron matrix 
$^\chi\fett{G}({}^\chi\fett{C})$ which, in HF theory, consists of a 
Coulomb matrix $^\chi\fett{J}({}^\chi\fett{C})$ and an exchange matrix $^\chi\fett{K}({}^\chi\fett{C})$,
\begin{equation}
 {}^{\chi}\!\fett{F}({}^\chi {\fett{C}}) = {}^{\chi}\!\fett{H} + {}^{\chi}\!\fett{G}({}^\chi {\fett{C}}) = 
{}^{\chi}\!\fett{H} + {}^{\chi}\!\fett{J}({}^\chi {\fett{C}}) + {}^{\chi}\!\fett{K}({}^\chi {\fett{C}}).
\end{equation}
After reaching self-consistency, the total electronic HF energy  of  the  system  is  calculated  
from  ${}^{\chi}\!{\fett{P}}({}^\chi {\fett{C}})$, ${}^\chi {\fett{F}}({}^\chi {\fett{C}})$, 
and the nucleus-nucleus repulsion energy $V$,
\begin{equation}
\label{eq:ehf}
\begin{split}
E_{\text{el}}^{\text{HF}}(^\phi\fett{C}) &= 
\frac{1}{2} \sum_{\mu\nu}^M 
\;{}^{\chi}\!{P}_{\nu\mu}({}^\chi {\fett{C}}) \left( 2\; {}^\chi\!{H}_{\mu\nu}+
{}^\chi\!{G}_{\mu\nu}({}^\chi {\fett{C}}) \right)  + V. \\
\end{split}
\end{equation}

We need to introduce a second basis, the symmetrically orthogonalized \cite{Lowdin1970}
 basis $\fett{\phi}$, to discuss the NDDO approximation.
The symmetrically orthogonalized basis functions $\fett{\phi}=\{\phi_\mu\}$
and the locally orthogonal basis functions $\fett{\chi}=\{\chi^I_\mu\}$  are related through 
\begin{equation}
 \label{eq:oao}
  {\phi_\nu} = \sum_{\mu=1}^M (\;{}^{\chi}\!{S}^{-\frac{1}{2}})_{\mu\nu}\ 
\chi_\mu^I.
\end{equation}
Consequently, we can calculate the ERIs in the $\phi$-basis ($^\phi$ERIs) by a transformation 
involving the 
$^\chi$ERIs, 
\begin{equation}
\label{eq:exactTransformation}
 \left<\phi_\mu\phi_\nu | \phi_\lambda \phi_\sigma\right>
= 
\sum_{\mu'\nu'\lambda'\sigma'}^M
({}^\chi {S}^{-\frac{1}{2}})_{\mu\mu'}
({}^\chi {S}^{-\frac{1}{2}})_{\nu\nu'}
\left<\chi_{\mu'}^I \chi_{\nu'}^J | \chi_{\lambda'}^K \chi_{\sigma'}^L\right>
({}^\chi {S}^{-\frac{1}{2}})_{\lambda\lambda'}
({}^\chi {S}^{-\frac{1}{2}})_{\sigma\sigma'}.
\end{equation}
This is formally a 4-index transformation which scales as $\mathcal{O}(M^5)$. \cite{Helgaker2014}

\subsection{Definition of the Approximation}

\begin{figure}[ht] 
 \centering
 \includegraphics[width=\textwidth]{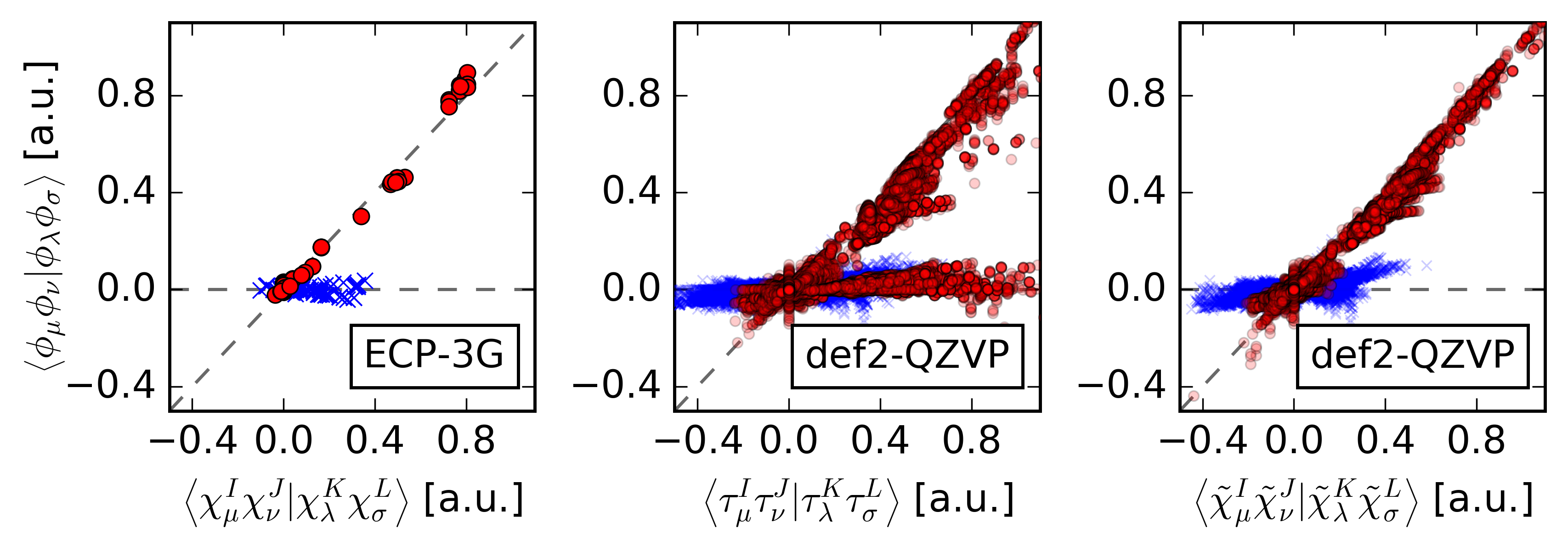}
 \caption{$^\phi$ERIs (y-axis), $^\chi$ERIs (x-axis, left), 
$^\tau$ERIs (x-axis, middle), and $^\chi$ERIs after local orthogonalization (x-axis, right) 
 for the water molecule in an ECP-3G \cite{Stevens1984,Kolb1993a} basis set (left) and in a def2-QZVP 
\cite{Weigend2005} basis set (middle and right).
The NDDO approximation holds if the red circles ($I=J$ and $K=L$) 
lie on the diagonal gray dashed line and if the blue crosses ($I\ne J$ or $K\ne L$) lie on 
the horizontal dashed line.
}
\label{fig:nddotau}
\end{figure}

The NDDO approximation provides a recipe for how to estimate 
$^\phi$ERIs based on a small number of $^\chi$ERIs, \cite{Pople1965} 
\begin{equation}
\label{eq:nddo}
\left<\phi_\mu \phi_\nu | \phi_\lambda \phi_\sigma\right>
\approx \delta_{IJ} \delta_{KL} \left<\chi^I_\mu \chi_\nu^J| \chi_\lambda^K \chi_\sigma^L\right>.
\end{equation}
As a consequence, the formal scaling of the $^\phi$ERI evaluation step is reduced from 
$\mathcal{O}(M^5)$ 
to $\mathcal{O}(M^2)$.
It is not immediately obvious why Eq.~(\ref{eq:nddo}) should hold true, especially 
 in view of Eq.~(\ref{eq:exactTransformation}), but numerical data supports it. 
\cite{Cook1967,Roby1969,Sustmann1969,Gray1970,
Roby1971, Brown1971,Roby1972,Brown1973,Birner1974,Chandler1980,Duke1981, Weinhold1988,Koch2014,
Neymeyr1995a,Neymeyr1995c,Neymeyr1995d,
Neymeyr1995,
Neymeyr1995b,Tu2003} 
The NDDO approximation, Eq.~(\ref{eq:nddo}), 
contains two central statements which are illustrated in Figure~\ref{fig:nddotau}
at the example of water.
First, the NDDO approximation states that a $^\phi$ERI will be similar to a 
$^\chi$ERI if $\chi^I_\mu$ and $\chi_\nu^J$ are centered on the same atom ($I=J$) and 
$\chi_\lambda^K$ and $\chi_\sigma^L$ are centered on the same atom ($K=L$), 
i.e., $\delta_{IJ}\delta_{KL}=1$,
\begin{equation}
 \left<\phi_\mu \phi_\nu | \phi_\lambda \phi_\sigma\right>
\approx \left<\chi^I_\mu \chi_\nu^I| \chi_\lambda^K \chi_\sigma^K\right>.
\end{equation}
We can see that this approximation is valid for the water example because the red circles in Figure~\ref{fig:nddotau} (left)  
are located close to the diagonal dashed line.
Second, the NDDO approximation states that a $^\phi$ERI will be zero if  
$\chi^I_\mu$ and $\chi_\nu^J$ are not centered on the same atom ($I\ne J$) or if  
$\chi_\lambda^K$ and $\chi_\sigma^L$ are not centered on the same atom  ($K\ne L$), 
i.e., $\delta_{IJ}\delta_{KL}=0$.
Consequently, the blue crosses in Figure~\ref{fig:nddotau} have to lie close to the horizontal dashed lines 
($\left<\phi_\mu \phi_\nu | \phi_\lambda \phi_\sigma\right>=0$) 
for the NDDO approximation to be reliable. 
If a blue cross does not lie close to this horizontal dashed line, it will indicate that the 
NDDO approximation does not hold. 
We examine the errors that Eq.~(\ref{eq:nddo})
introduces in $^\phi$ERIs in detail.

Throughout this work, we denote the error which arises from the application 
of Eq.~(\ref{eq:nddo}) instead of Eq.~(\ref{eq:exactTransformation}) by $\mathcal{E}$. 
The superscript to $\mathcal{E}$ indicates which 
quantity is affected; additional
specifications are then given as subscripts.
For example, the error introduced by the NDDO approximation for the $^\phi$ERI 
$\left<\phi_\mu\phi_\nu | \phi_\lambda \phi_\sigma\right>$
is denoted as $\mathcal{E}^{^\phi\text{ERI}}_{\mu\nu\lambda\sigma}$.
We define $\mathcal{E}^{^\phi\text{ERI}}_{\mu\nu\lambda\sigma}$ as 
 the deviation of $\delta_{IJ}\delta_{KL}
\left<\chi_\mu^I \chi_\nu^J | \chi_\lambda^K \chi_\sigma^L\right>$
 from the analytical value of $\left<\phi_\mu\phi_\nu | \phi_\lambda \phi_\sigma\right>$,
\begin{equation}
 \mathcal{E}^{^\phi\text{ERI}}_{\mu\nu\lambda\sigma} = 
\left<\phi_\mu\phi_\nu | \phi_\lambda \phi_\sigma\right>-\delta_{IJ}\delta_{KL}
\left<\chi_\mu^I \chi_\nu^J | \chi_\lambda^K \chi_\sigma^L\right>.
\end{equation}
Obviously, $M^4$ different errors $\mathcal{E}^{^\phi\text{ERI}}_{\mu\nu\lambda\sigma}$ 
need to be accounted for.  

\begin{figure}[ht]
 \centering
 \includegraphics[width=.5\textwidth]{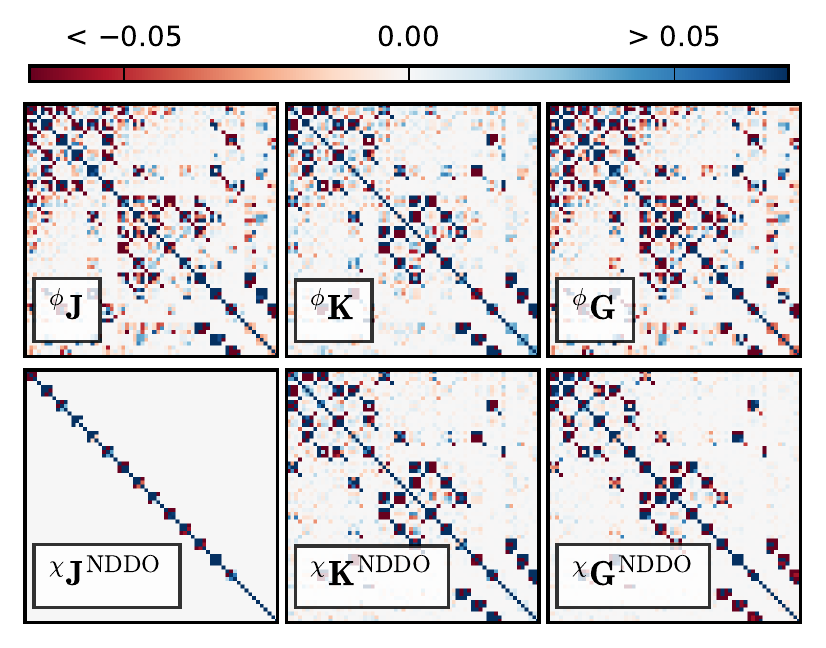}
 \caption{Graphical representation of $^\phi\!\fett{J}(^\phi\fett{C})$, $^\phi\fett{K}(^\phi\fett{C})$,
and $^\phi\fett{G}(^\phi\fett{C})$ (left to right in upper panel) and 
of $^\chi\!\fett{J}^{\text{NDDO}}(^\phi\fett{C})$, $^\chi\fett{K}^{\text{NDDO}}(^\phi\fett{C})$,
and $^\chi\fett{G}^{\text{NDDO}}(^\phi\fett{C})$ (left to right in lower panel) of the 
caffeine molecule (ECP-3G basis set). 
The matrices were evaluated with a density matrix determined from a fully converged HF calculation 
yielding $^\phi\fett{C}$. 
They are colored according to their values ranging from $-0.05$ a.u.\ (red) to 
zero (white) to $0.05$ a.u.\ (blue).
}
\label{fig:nddojkg}
\end{figure}

We then consider the effect of erroneous $^\phi$ERIs on the HF energy when 
 the self-consistent solution $^\phi\fett{C}$ obtained from an exact HF calculation
is applied. 
The errors in the $^\phi$ERIs affect the Coulomb matrix elements,  
\begin{equation}
\label{eq:Jmatrix_phi}
\begin{split}
 \;{}^{\phi}\!{J}_{\mu\nu}({}^\phi {\fett{C}}) 
\approx \;{}^{\chi}\!{J}^{\text{NDDO}}_{\mu\nu}({}^\phi {\fett{C}}) &= 
\sum_{\lambda\sigma}^M \;{}^{\phi}\!{P}_{\lambda\sigma}({}^\phi {\fett{C}})
\delta_{IJ} \delta_{KL}
\left<\chi_\mu^I\chi_\nu^J|\chi^K_\lambda\chi^L_\sigma\right>, \\
\end{split}
\end{equation}
and the exchange matrix elements, 
\begin{equation}
\label{eq:Kmatrix_phi}
\begin{split}
 \;{}^{\phi}\!{K}_{\mu\nu}({}^\phi {\fett{C}}) 
\approx \;{}^{\chi}\!{K}^{\text{NDDO}}_{\mu\nu}({}^\phi {\fett{C}}) =  -\frac{1}{2}
\sum_{\lambda\sigma}^M  \;{}^{\phi}\!{P}_{\lambda\sigma}({}^\phi {\fett{C}})               
\delta_{IL} \delta_{JK}
\left<\chi^I_\mu\chi^L_\sigma|\chi^K_\lambda\chi^J_\nu\right>.
\end{split}
\end{equation}
Interestingly, the matrix element ${}^{\phi}\!{J}_{\mu\nu}$ will always be exactly zero if  
$\chi_\mu^I$ and $\chi_\nu^J$ are centered on different atoms ($I\ne J$)
irrespective of the number of atoms on which $\chi_\lambda^K$ and $\chi_\sigma^L$ are centered 
(see Figure~\ref{fig:nddojkg}). 
By contrast,  ${}^{\phi}\!{K}_{\mu\nu}$ will not be strictly zero in this case 
 (see also Figure~\ref{fig:nddojkg}).
By affecting the Coulomb and exchange matrices, the NDDO approximation will introduce an 
error compared to $E^{\text{HF}}_{\text{el}}$ which we denote 
by $\mathcal{E}^{\text{HF}}=\mathcal{E}^{E^{\text{HF}}_{\text{el}}}$,
\begin{equation}
\begin{split}
\mathcal{E}^{\text{HF}} =& E_{\text{el}}^{\text{HF}}({}^\phi {\fett{C}}) - 
E_{\text{el}}^{\text{HF-NDDO}}({}^\phi {\fett{C}}), \\
=& \frac{1}{2}\sum_{\mu\nu\lambda\sigma}^M {}^\phi {P}_{\nu\mu}({}^\phi {\fett{C}})  
{}^\phi {P}_{\lambda\sigma}({}^\phi {\fett{C}})
\left[ \mathcal{E}^{^\phi\text{ERI}}_{\mu\nu\lambda\sigma} - \frac{1}{2} 
\mathcal{E}_{\mu\sigma\lambda\nu}^{^\phi\text{ERI}} \right],\\
\end{split}
\end{equation}
in the closed-shell case.

If we apply the NDDO approximation, we, however, must iterate to 
self consistency. The self-consistently obtained $^\chi\fett{C}^{\text{NDDO}}$ 
will likely not be the same as $^\phi\fett{C}$.
Hence, another error arises from the NDDO approximation by introducing 
errors in other quantities during the self-consistent-field (SCF) cycles, i.e., in the coefficient matrix and 
in the matrix of orbital energies. 
We denote this error by $\mathcal{G}$ and again 
indicate by a superscript which 
quantity is affected by the error (and give additional
specifications as subscripts). By contrast, 
$\mathcal{E}$ denotes the error that is obtained when applying $^\phi\fett{C}$.
The difference of the two self-consistent solutions produces 
$\mathcal{G}^{\text{HF}}=\mathcal{G}^{E^{\text{HF}}_{\text{el}}}$,
\begin{equation}
\label{eq:ghf}
\begin{split}
 \mathcal{G}^{\text{HF}} = & 
E^{\text{HF}}_{\text{el}}({}^\phi {\fett{C}}) - 
E^{\text{HF-NDDO}}_{\text{el}}({}^\chi{\fett{C}}^{\text{NDDO}}), \\
=& 
\sum_{\mu\nu}^M ({}^\phi {P}_{\nu\mu}({}^\phi {\fett{C}})-{}^\chi {P}^{\text{NDDO}}_{\nu\mu}({}^\chi{\fett{C}}^{\text{NDDO}})) 
\left<\phi_\mu|\hat{h}|\phi_\nu\right>  \\
& 
+ \ \frac{1}{2} \sum_{\mu\nu\lambda\sigma}^M 
\left(
{}^\phi {P}_{\nu\mu}({}^\phi {\fett{C}}) {}^\phi {P}_{\lambda\sigma}({}^\phi {\fett{C}}) 
\left[ 
\left<\phi_\mu\phi_\nu|\phi_\lambda\phi_\sigma\right> 
                 - \frac{1}{2}  
\left<\phi_\mu\phi_\sigma|\phi_\lambda\phi_\nu\right>
\right] \right. \\
& \left. - 
{}^\chi {P}^{\text{NDDO}}_{\nu\mu}({}^\chi{\fett{C}}^{\text{NDDO}}) 
{}^\chi {P}^{\text{NDDO}}_{\lambda\sigma}({}^\chi{\fett{C}}^{\text{NDDO}}) 
\left[ 
\delta_{IJ}\delta_{KL} 
\left<\chi^I_\mu\chi^J_\nu|\chi^K_\lambda\chi^L_\sigma\right>
                 \right. \right. \\  
& \left. \left. - \frac{1}{2} \delta_{IL} \delta_{JK} 
\left<\chi^I_\mu\chi^L_\sigma|\chi^K_\lambda\chi^J_\nu\right>
\right] \right).\\
\end{split}
\end{equation}

The electronic HF energy does, by definition, not contain effects from electron 
correlation. \cite{Helgaker2014}
Various electronic structure methods are available for calculating 
correlation energies. \cite{Helgaker2014}
The prevalent single- and multi-reference methods require the calculation of  
 ERIs in the molecular orbital basis $\fett{\psi}$ ($^\psi$ERIs).
These $^\psi$ERIs are obtained 
through a 4-index transformation of the $^\phi$ERIs (or the $^\chi$ERIs),
\begin{equation}
\label{eq:erimo}
 \left<\psi_i\psi_j|\psi_k\psi_l\right> = 
\sum_{\mu\nu\lambda\sigma}^M {}^\phi C_{\mu i} {}^\phi C_{\nu j}
\left<\phi_\mu\phi_\nu|\phi_\lambda\phi_\sigma\right> 
{}^\phi C_{\lambda k} {}^\phi C_{\sigma l}.  
\end{equation}
This 4-index transformation is similar to the 4-index transformation with which 
the $^\phi$ERIs are determined from the $^\chi$ERIs (Eq.~(\ref{eq:exactTransformation})).
When applying Eq.~(\ref{eq:nddo}), we may approximate the $^\psi$ERIs as,
\begin{equation}
\label{eq:erimo2}
 \left<\psi_i\psi_j|\psi_k\psi_l\right> \approx 
\sum_{\mu\nu\lambda\sigma}^M {}^\phi C_{\mu i} {}^\phi C_{\nu j}\;
\delta_{IJ} \delta_{KL} 
\left<\chi_\mu^I\chi_\nu^J|\chi_\lambda^K\chi_\sigma^L\right>  
{}^\phi C_{\lambda k} {}^\phi C_{\sigma l}.  
\end{equation}
The formal scaling of the $^\psi$ERI evaluation step is therefore reduced from $\mathcal{O}(M^5)$ 
to $\mathcal{O}(M^2)$ scaling which comes at the price of an error in the $M^4$ $^\psi$ERIs, 
$\mathcal{E}^{{}^\psi\text{ERI}}_{ijkl}$,
\begin{equation}
\mathcal{E}^{^\psi\text{ERI}}_{ijkl} = 
\sum_{\mu\nu\lambda\sigma}^M  {}^\phi C_{\mu i} {}^\phi C_{\nu j} \;
\mathcal{E}^{^\phi\text{ERI}}_{\mu\nu\lambda\sigma} \; {}^\phi C_{\lambda k} {}^\phi C_{\sigma l}.
\end{equation}
If we determine the coefficient matrix in a self-consistent field procedure, we will introduce an additional 
error $\mathcal{G}^{^\psi\text{ERI}}_{ijkl}$
from applying a different coefficient matrix, 
\begin{equation}
\begin{split}
 \mathcal{G}^{^\psi\text{ERI}}_{ijkl} =& \sum_{\mu\nu\lambda\sigma}^M 
 \left ( {}^\phi C_{\mu i} {}^\phi C_{\nu j}
\left<\phi_\mu\phi_\nu|\phi_\lambda\phi_\sigma\right> 
{}^\phi C_{\lambda k} {}^\phi C_{\sigma l} \right. \\ 
& \left. -
{}^\chi C^{\text{NDDO}}_{\mu i} {}^\chi C^{\text{NDDO}}_{\nu j} \;
\delta_{IJ} \delta_{KL} 
\left<\chi_\mu^I\chi_\nu^J|\chi_\lambda^K\chi_\sigma^L\right>  
{}^\chi C^{\text{NDDO}}_{\lambda k} {}^\chi C^{\text{NDDO}}_{\sigma l}
\right).
\end{split}
\end{equation}
In this work, we demonstrate how $\mathcal{E}^{^\psi\text{ERI}}_{ijkl}$ and 
$\mathcal{G}^{^\psi\text{ERI}}_{ijkl}$ affect the MP2 correlation energies.
We denote the total MP2 energies as 
$E^\text{MP2}_\text{el}$ which is then given as 
\begin{equation}
\label{eq:defmp2}
 E^\text{MP2}_\text{el} = E^\text{(0)}_\text{el} + E^\text{(1)}_\text{el} + E^\text{(2)}_\text{el} 
= E^\text{HF}_\text{el} + E^\text{(2)}_\text{el},
\end{equation}
where $E^\text{(0)}_\text{el}$, $E^\text{(1)}_\text{el}$, and $E^\text{(2)}_\text{el}$ denote the 
low-order perturbation-theory contributions. 
We quantify $\mathcal{E}^{\text{(2)}}=\mathcal{E}^{E^\text{(2)}_\text{el}}$ and 
$\mathcal{G}^{\text{(2)}}=\mathcal{G}^{E^\text{(2)}_\text{el}}$, respectively, 
by subtracting $E^\text{(2)}_\text{el}$ obtained when invoking the NDDO approximation from the 
exact  $E^\text{(2)}_\text{el}$. 

\subsection{Extension to Conventional Basis Sets}

\begin{figure}[ht]
 \centering
 \includegraphics[width=.4\textwidth]{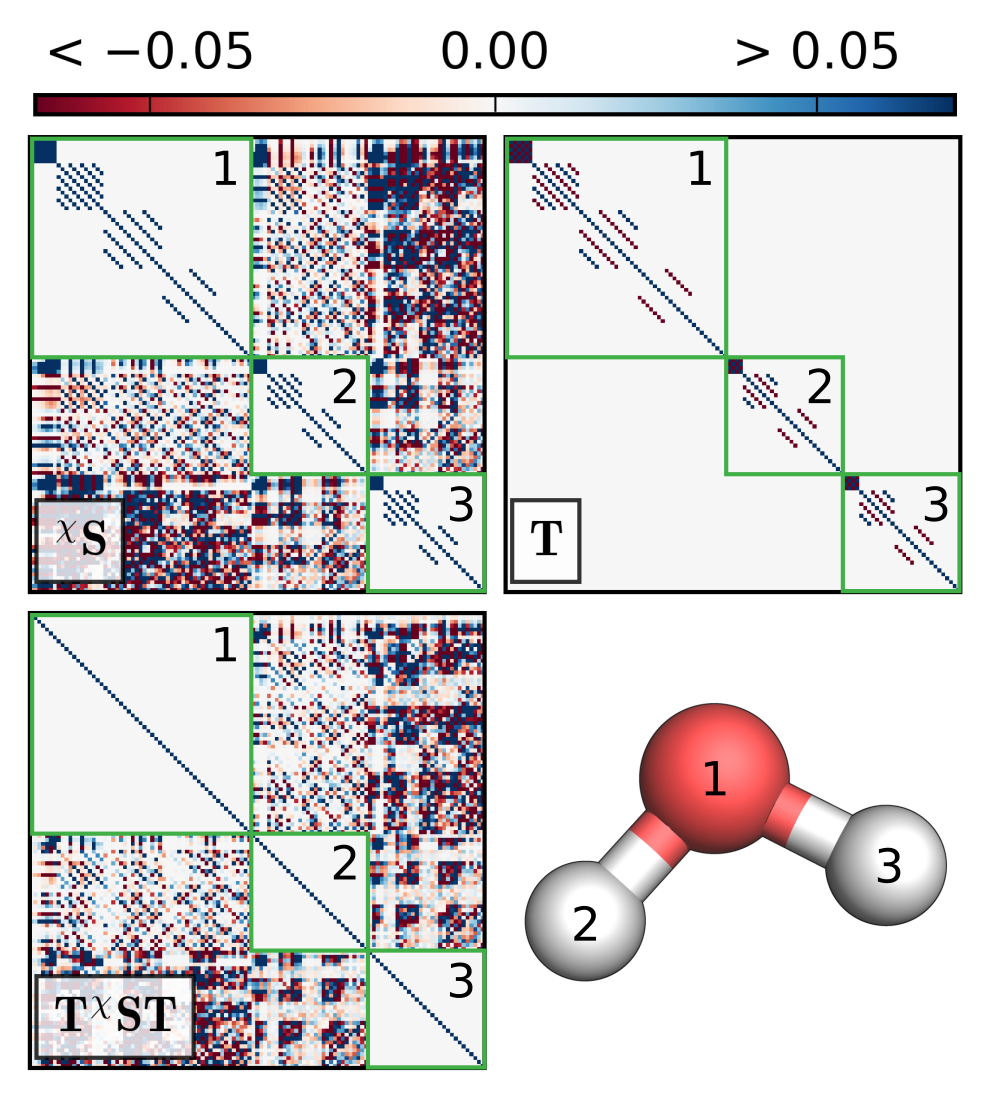}
 \caption{
Graphical representation of $^\chi\fett{S}$, $\fett{T}$, and $\fett{T}^\chi\fett{S}\fett{T}$
for a water molecule (def2-QZVP basis set).
The entries according to their values ranging from $-0.05$ (red) to 
zero (white) to $0.05$ (blue). The blocks for which the respective basis functions are centered on the 
same atom are highlighted in green.
}
\label{fig:TS}
\end{figure}

By definition, the $\chi$-basis fulfills the condition that it is locally 
orthogonal. Ordinary basis sets are, in general, not locally orthogonal which we illustrate 
in Figure~\ref{fig:TS} for the example of water and a def2-QZVP basis set.
If the basis set would be locally orthogonal, we would not have any off-diagonal entries in 
the green boxes in Figure~\ref{fig:TS}.
We denote such ordinary Gaussian-type basis functions by $\tau_\mu^I$
($\mu$-th basis function of type $\tau$ centered on atom $I$).
The NDDO approximation is not straightforwardly 
applicable for an arbitrary $\tau$-basis which is also illustrated in Figure~\ref{fig:nddotau} (middle).
When we apply an ordinary basis set, such as the def2-QZVP basis set \cite{Weigend2005}, 
many of the red circles are not 
located close to the diagonal dashed lines anymore, i.e., 
\begin{equation}
\label{eq:notnddo}
 \left< \phi_\mu \phi_\nu | \phi_\lambda \phi_\sigma\right>
\not\approx \left<\tau_\mu^I \tau_\nu^I | 
\tau_\lambda^K \tau_\sigma^K\right>.
\end{equation}
To cure this problem, we propose to transform $\{\tau_\mu^I\}$ to a locally orthogonal basis 
$\{\chi_\mu^I\}$ by the application of the transformation matrix $\fett{T}$ (see Figure~\ref{fig:TS}),
\begin{equation}
 \label{eq:tao}
  \begin{split}
  T_{\mu\nu} &= \delta_{IJ} ({}^\chi {S}^{-\frac{1}{2}})_{\mu\nu}, \\
  \end{split}
\end{equation}
 so that 
\begin{equation}
 \label{eq:tao2}
  \begin{split}
  {\tilde{\chi}_\nu^J} &= \sum_{\mu=1}^M {T}_{\mu\nu}
\tau_\mu^I, \\
  \end{split}
\end{equation}
where the tilde indicates that the local orthogonal basis function $\tilde{\chi}$ was obtained by means of 
a transformation of a $\tau$-basis. 
Figure~\ref{fig:TS} illustrates that the basis $\tilde{\chi}$ for the water example 
is locally orthogonal which is evident from the fact that there are no nonzero 
off-diagonal elements in the green boxes for $\fett{T}^\chi\fett{S}\fett{T}$.  
Figure~\ref{fig:nddotau} (right) shows that all red circles are  located close to the diagonal dashed line
 after local orthogonalization, i.e., the validity of the NDDO approximation has been restored.

\section{Analysis of the NDDO Approximation for Molecules in a $\chi$-Basis}

For the first part of our analysis of the NDDO approximation, 
we applied the ECP-3G basis set. \cite{Stevens1984,Kolb1993a} 
We selected the ECP-3G basis set because its basis functions 
 form a $\chi$-basis and because it is applied (in slightly modified forms) 
in the OM1, OM2, and OM3 models. \cite{Kolb1993a,Weber2000a,Dral2016b} 
The ECP-3G basis set specifies one $s$-type basis function for hydrogen and 
one $s$-type and three $p$-type basis functions for carbon, nitrogen, oxygen, and fluorine.
We consider four different sets of molecular structures for this part of the analysis:
(A) We first analyze the NDDO approximation on the simplest possible model 
system which is a dihydrogen molecule H$_1$---H$_2$ with an interatomic distance $R_{12}$. 
This is the simplest possible model system because NDDO 
is no approximation for isolated atoms
(where $\fett{\phi}=\fett{\chi}$) and for systems with only one electron (e.g., H$_2^+$).
(B) We assemble a series of linear alkane chains 
C$_x$H$_{2x+2},\ x=1, 2,..., 15$ to study trends with an increasing molecular size. 
(C) We randomly select a subset of 5000 molecules of the QM9 data set 
\cite{Ruddigkeit2012,Ramakrishnan2014} which allows us to examine the NDDO approximation 
for a variety of equilibrium structures of molecules composed of 
H, C, N, O, and F. 
(D) We choose three reaction trajectories, 
a Diels--Alder reaction between butadiene and ethylene yielding cyclohexene (reaction A), 
the decomposition of azobisisobutyronitrile  (reaction E), and the 
elimination of CO$_2$ from the benzoyl radical 
(reaction F) which we published in previous work. \cite{Muhlbach2016}

\subsection{Effect on Electron-Electron Repulsion Integrals}

\begin{figure}[ht] 
 \centering
 \includegraphics[width=\textwidth]{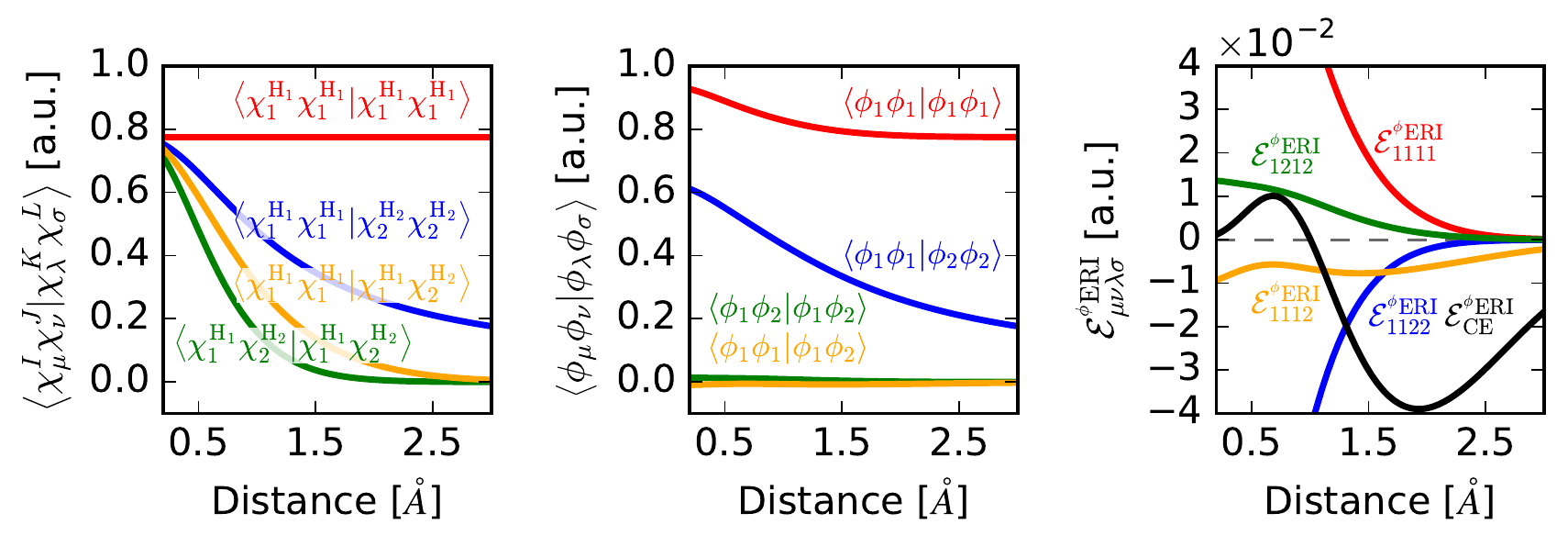}
 \caption{Dependence of the $^\chi$ERIs (left), the $^\phi$ERIs (middle), and 
the error in the $^\phi$ERIs (right) on $R_{12}$ in an H$_2$ molecule described by an ECP-3G basis. 
If the NDDO approximation would hold true, the error in the $^\phi$ERIs would be zero for every 
distance.
}
\label{fig:nddoh2_a}
\end{figure}

For the simplest possible model system, H$_2$, $2^4=16$ ERIs arise for a given $R_{12}$. 
Due to symmetry relations, only four of these 16 values are different, \cite{Szabo2012}
so that it suffices to discuss $\mathcal{E}_{1111}^{^\phi\text{ERI}}$, 
$\mathcal{E}_{1122}^{^\phi\text{ERI}}$, 
$\mathcal{E}_{1212}^{^\phi\text{ERI}}$, and $\mathcal{E}_{1112}^{^\phi\text{ERI}}$
(see Figure~\ref{fig:nddoh2_a}).
If the NDDO approximation would be valid, $\left<\chi^{\text{H}_1}_1\chi^{\text{H}_1}_1 | \chi^{\text{H}_1}_1 
\chi^{\text{H}_1}_1\right>$
would be similar to 
$\left<\phi_1\phi_1 | \phi_1 \phi_1\right>$ (red lines in left and middle panels of Figure~\ref{fig:nddoh2_a},
respectively)
and 
$\left<\chi^{\text{H}_1}_1\chi^{\text{H}_1}_1 | \chi^{\text{H}_2}_2 
\chi^{\text{H}_2}_2\right>$
would be similar to 
$\left<\phi_1\phi_1 | \phi_2 \phi_2\right>$ (blue lines in left and middle panels of Figure~\ref{fig:nddoh2_a},
respectively). The right panel of Figure~\ref{fig:nddoh2_a} visualizes the resulting error in these 
two $^\phi$ERIs. 
The error introduced by the NDDO approximation in $\left<\phi_1\phi_1 | \phi_1 \phi_1\right>$
and $\left<\phi_1\phi_1 | \phi_2 \phi_2\right>$ is large ($> 0.02$ a.u.) for 
$R_{12} < 2.0$ \AA. Only if the overlap between  $\chi_1^{\text{H}_1}$
and $\chi_2^{\text{H}_2}$ is small at large $R_{12}$, i.e., where the $\chi$-basis becomes a 
$\phi$-basis, $|\mathcal{E}_{1111}^{^\phi\text{ERI}}|$ and 
$|\mathcal{E}_{1122}^{^\phi\text{ERI}}|$ will be small. 
The $^\phi$ERIs $\left<\phi_1\phi_2 | \phi_1 \phi_2\right>$ and 
$\left<\phi_1\phi_1 | \phi_1 \phi_2\right>$
are assumed to be zero in the NDDO approximation (green and orange lines in the 
middle panel of Figure~\ref{fig:nddoh2_a}) which seems to be a good approximation. 
Figure~\ref{fig:nddoh2_a} clearly illustrates that the corresponding $^\chi$ERIs are not zero.
 For $R_{12} < 1.6$ \AA, 
$|\mathcal{E}_{1212}^{^\phi\text{ERI}}|$ and $|\mathcal{E}_{1112}^{^\phi\text{ERI}}|$ are smaller 
than $|\mathcal{E}_{1111}^{^\phi\text{ERI}}|$ and $|\mathcal{E}_{1122}^{^\phi\text{ERI}}|$.
For $R_{12}$ larger than 1.6 \AA\ , $\mathcal{E}_{1112}^{^\phi\text{ERI}}$ is the 
largest individual error in a $^\phi$ERI. 
This gives rise to a large cumulative error (CE) $\mathcal{E}^{^\phi\text{ERI}}_\text{CE}$, 
\begin{equation}
\label{eq:ce}
 \mathcal{E}^{^\phi\text{ERI}}_\text{CE} =\sum_{\mu\nu\lambda\sigma}^M 
\mathcal{E}^{^\phi\text{ERI}}_{\mu\nu\lambda\sigma}, 
\end{equation} 
 for $1.6 $\ \AA\ $< R_{12} < 3.0 $ \AA.
For  $1.6 $\ \AA\ $< R_{12} < 3.0 $ \AA, we find a significant error in at least one of the $^\phi$ERIs. 

We also encounter nonnegligible $\mathcal{E}_{\mu\nu\lambda\sigma}^{^\phi\text{ERI}}$ for all structures in 
the sets of molecules B, C, and 
D. The largest absolute error of a $^\phi$ERI in a given molecule 
is between 0.10 a.u.\ and 0.23 a.u. It is already obvious at this point 
that only an efficient error cancellation may yield useful
observables based on these erroneous $^\phi$ERIs (and on density matrices obtained with them in SCF
procedures).

\begin{figure}[ht]
 \centering
 \includegraphics[width=\textwidth]{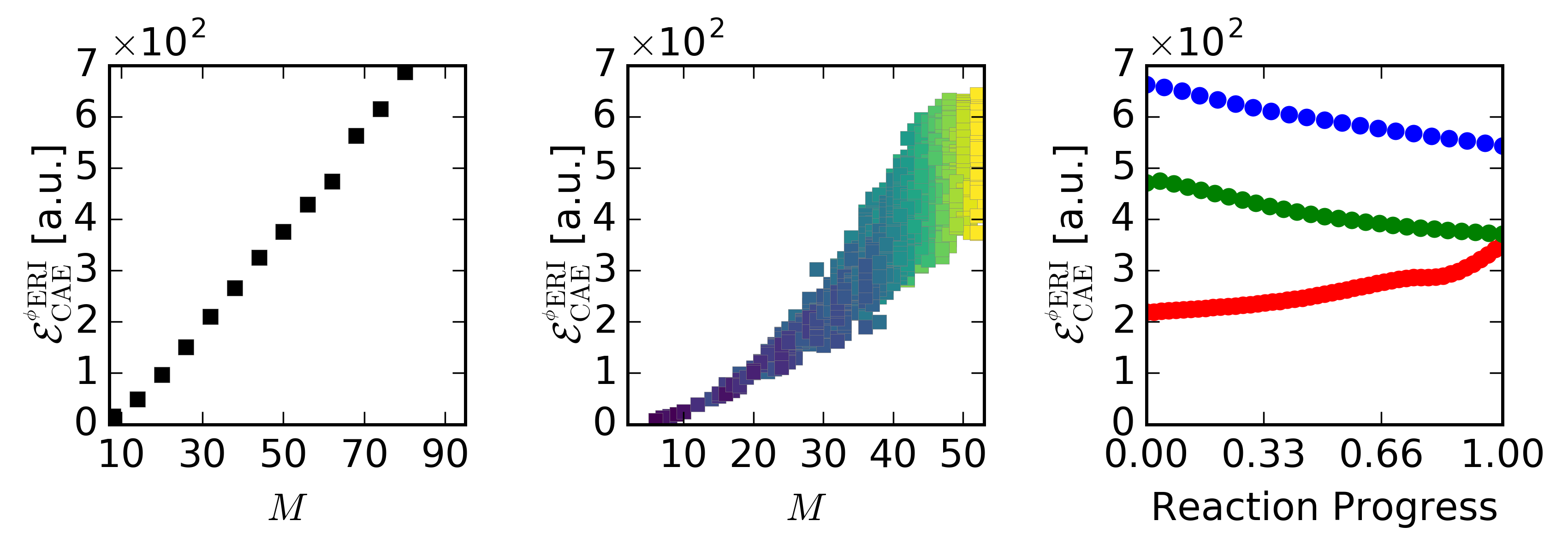}
 \caption{Dependence of the cumulative absolute error 
$\mathcal{E}_\text{CAE}^{^\phi\text{ERI}}$ (Eq.~(\ref{eq:cae})) 
on the number of basis functions $M$ 
for the molecule sets B (left), C (middle), and D (right). 
The entries for molecule set C are colored according to the 
number of atoms from purple (3 atoms) over blue (12 atoms) 
to yellow (25 atoms) and the ones for molecule set C according to the reaction:
Reaction A (34 basis functions; red circles), reaction E (60 basis functions; blue circles), and 
reaction F (41 basis functions; green circles). All calculations were carried out with 
the ECP-3G basis.}
\label{fig:errorAccumulation}
\end{figure}
Figure~\ref{fig:errorAccumulation} shows that the cumulative absolute error (CAE) in the $^\phi$ERIs, 
\begin{equation}
\label{eq:cae}
 \mathcal{E}^{^\phi\text{ERI}}_\text{CAE} =\sum_{\mu\nu\lambda\sigma}^M 
|\mathcal{E}^{^\phi\text{ERI}}_{\mu\nu\lambda\sigma}|, 
\end{equation} 
grows roughly linearly with the number of basis functions for molecule set B  
(Figure~\ref{fig:errorAccumulation}). 
We observe a linear growth not only in the overall cumulative absolute error, 
but also in individual contributions to it when we break down the 
corresponding $^\chi$ERIs in one-, two-, three-, and four-center $^\chi$ERIs (see Figure~S1 in 
the Supporting Information).
Figure~\ref{fig:errorAccumulation} shows that  
 $\mathcal{E}_\text{CAE}^{^\phi\text{ERI}}$
also grows approximately linearly with the number of basis functions for C
and that the spread of the individual 
$\mathcal{E}_\text{CAE}^{^\phi\text{ERI}}$ is 
large. 
In agreement with previous studies, \cite{Neymeyr1995,Neymeyr1995a,Neymeyr1995b,
Neymeyr1995c,Neymeyr1995d,Chandler1980} we find that the assumption 
that the $^\phi$ERIs corresponding to 
$\left<\chi_\mu^I \chi_\nu^I | \chi_\lambda^J \chi_\sigma^K\right>$ for $\ I\ne J\ne K$ 
and 
$\left<\chi_\mu^I \chi_\nu^I | \chi_\lambda^I \chi_\sigma^J\right>$ for $\ I\ne J$  are zero 
is responsible for 60--65\% of the 
overall cumulative absolute error for the molecule sets B and C (see Figure~S1
 in the Supporting Information). 
Furthermore, the change of $\mathcal{E}_\text{CAE}^{^\phi\text{ERI}}$ with  
reaction progress for D in Figure~\ref{fig:errorAccumulation} 
shows that the cumulative absolute error in the $^\phi$ERIs 
crucially depends on the arrangement of the atomic nuclei, i.e. on the underlying
nuclear framework that generates the external potential. 
If the cumulative absolute error in the $^\phi$ERIs would not depend on the arrangement of the 
atomic nuclei, it would not change in the course of the reaction.

\subsection{Error Propagation: the Hartree--Fock Energy}

For H$_2$, $\mathcal{E}^{\text{HF}}$ depends linearly 
on the cumulative error of the $^\phi$ERIs because 
$^\phi\fett{C}$ can be determined analytically, \cite{Szabo2012} 
\begin{equation}
  \mathcal{E}^{\text{HF}} 
= \frac{1}{4} \mathcal{E}_\text{CE}^{^\phi\text{ERI}}
= \frac{1}{2} 
\mathcal{E}_{1111}^{^\phi\text{ERI}}+ \frac{1}{2} \mathcal{E}_{1122}^{^\phi\text{ERI}} +  
\mathcal{E}_{1212}^{^\phi\text{ERI}} + 2\;\mathcal{E}_{1112}^{^\phi\text{ERI}}.
\end{equation}
Previous results show that the NDDO approximation introduces an error of 
0.002 a.u.\ for H$_2$ with $R_{12} = 0.84$ \AA\ \cite{Koster1972, Tu2003}
which we can reproduce. In our detailed analysis however, we
also see that this value of $R_{12}$ falls into the region where 
$\mathcal{E}_\text{CE}^{^\phi\text{ERI}}$ is 
small due to a fortunate error cancellation of $\frac{1}{2}\mathcal{E}_{1111}^{^\phi\text{ERI}}$,
$\frac{1}{2} \mathcal{E}_{1122}^{^\phi\text{ERI}}$, 
$\mathcal{E}_{1212}^{^\phi\text{ERI}}$, and $2\;\mathcal{E}_{1112}^{^\phi\text{ERI}}$
(see right panel of Figure~\ref{fig:nddoh2_a}).
For larger or smaller $R_{12}$, we encounter larger $\mathcal{E}^{\text{HF}}$ because 
the errors in the $^\phi$ERIs do not cancel as effectively. 
\begin{figure}[ht!]
 \centering
 \includegraphics[width=\textwidth]{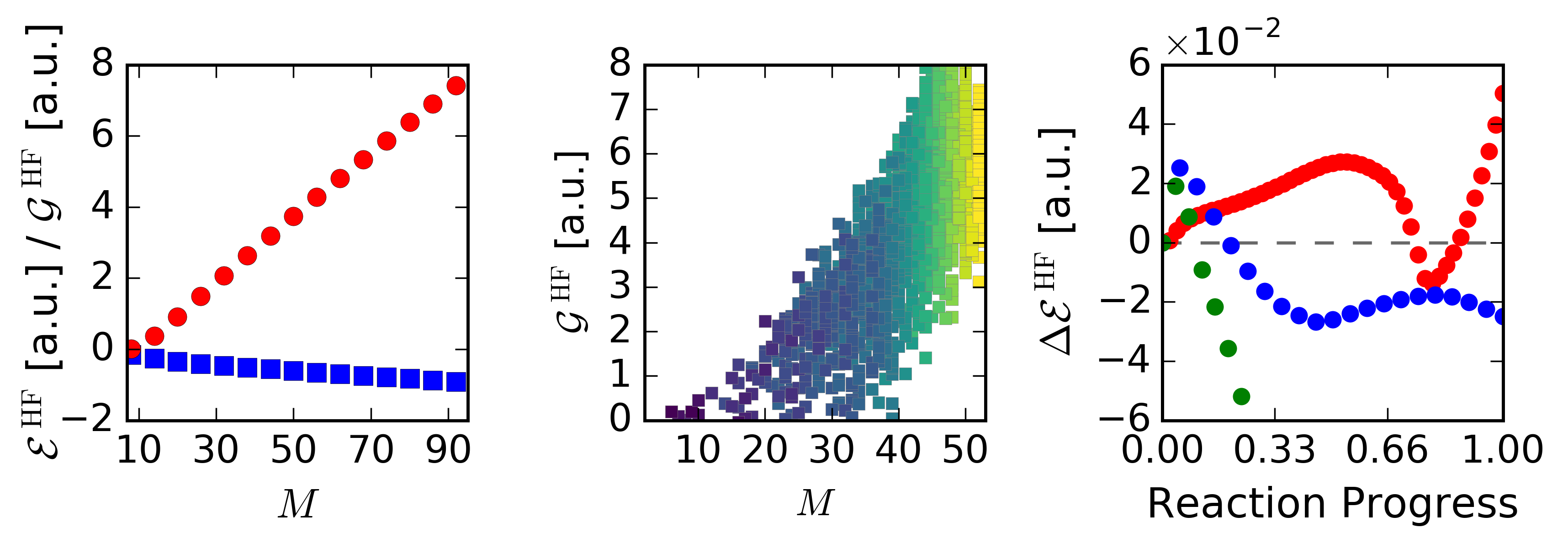}
 \caption{Left: Dependence of $\mathcal{E}^\text{HF}$ (blue squares) and $\mathcal{G}^\text{HF}$ 
(red circles) on the number of basis functions $M$ for molecules in B. 
Middle: Dependence of  $\mathcal{G}^\text{HF}$ on $M$ for molecules in 
C.
The entries for molecule set C are colored according to the 
number of atoms from purple (3 atoms) over blue (12 atoms) 
to yellow (25 atoms). 
Right: Change in $\mathcal{E}^\text{HF}$ with the reaction progress  
for reaction A (red circles), reaction E (blue circles), 
and reaction F (green circles). All calculations were carried out with 
an ECP-3G basis set.
}
\label{fig:errorAccumulation2}
\end{figure}

For the calculation of $E^{\text{HF}}$, the $^\phi$ERIs 
are contracted with elements of the density matrix. The elements of the density matrix are therefore a
central ingredient for an efficient error cancellation.
We see how differently the errors in the $^\phi$ERIs add up by comparing 
$\mathcal{E}^{\text{HF}}$ and $\mathcal{G}^{\text{HF}}$ for the molecule set 
B in Figure~\ref{fig:errorAccumulation2}.
Interestingly, $|\mathcal{E}^{\text{HF}}|$ and $\mathcal{G}^{\text{HF}}$ increase 
in an almost perfectly linear fashion with the number of basis functions. 
For a given number of basis functions,  $\mathcal{G}^{\text{HF}}$ 
is significantly larger than $|\mathcal{E}^{\text{HF}}|$ for molecule set B.
We also see a roughly linear increase of 
$\mathcal{G}^{\text{HF}}$  with the number of basis functions for the diverse organic molecules 
contained in set C. 
The errors in the HF energies also depend crucially on the arrangement of the atomic nuclei 
which we demonstrate in the right panel of Figure~\ref{fig:errorAccumulation2}. 
In line with previous studies, 
\cite{Sustmann1969, Tu2003,Chandler1980,Neymeyr1995,Neymeyr1995a,
 Neymeyr1995b,Neymeyr1995c,Neymeyr1995d}
we see that 
the NDDO approximation is a rather crude approximation. We 
discuss strategies to overcome this situation   
in Section~\ref{sec:improve}.

\subsection{Error Propagation: the MP2 Energy}
\label{subsec:mp2}

\begin{figure}[ht] 
 \centering
 \includegraphics[width=.85\textwidth]{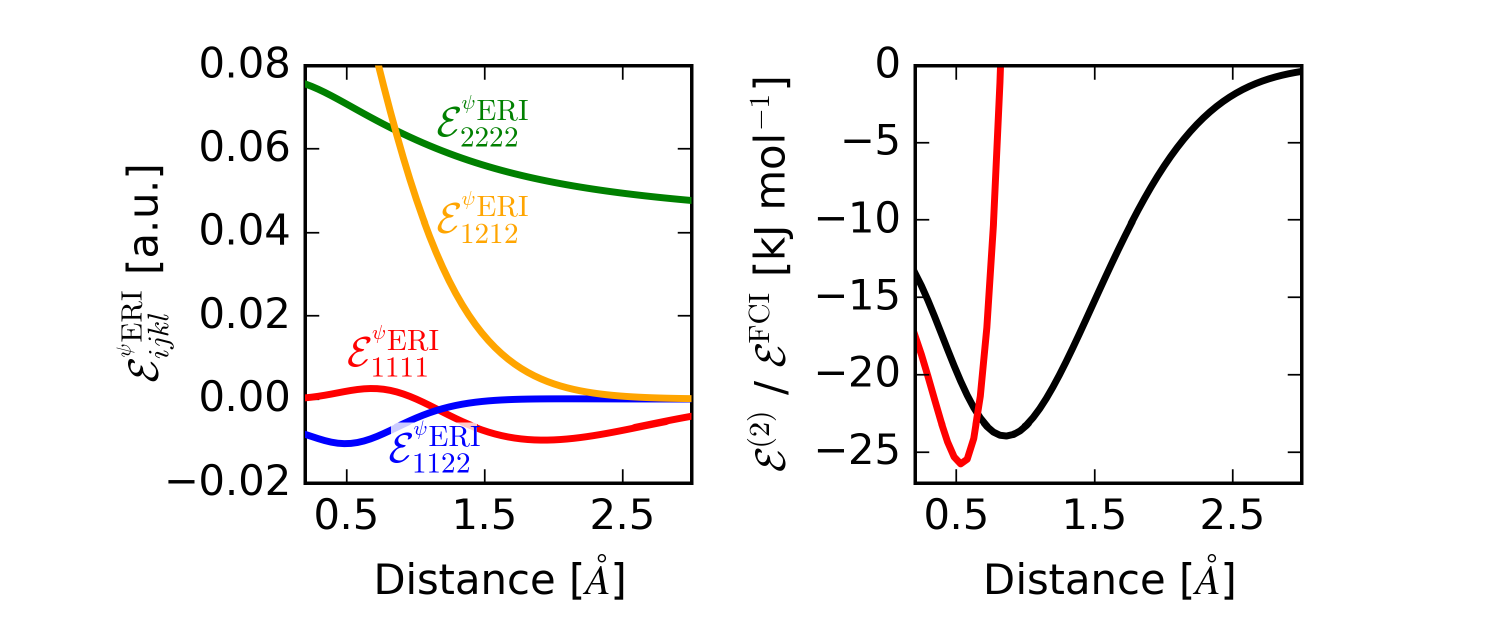}
 \caption{
Dependence of the error in selected $^\psi$ERIs (left),  
$\mathcal{E}^{\text{MP2}}$ (black line; right), and $\mathcal{E}^{\text{FCI}}$ (red line;  right)
on $R_{12}$ in the H$_2$ molecule described in an ECP-3G basis.
}
\label{fig:nddoh2_b}
\end{figure}

The error in the $^\phi$ERIs also propagates to the $^\psi$ERIs
(Eq.~(\ref{eq:erimo2})). 
We show the effect of the NDDO approximation on selected $^\psi$ERIs,
$\left<\psi_1\psi_1|\psi_1\psi_1\right>$,
$\left<\psi_1\psi_1|\psi_2\psi_2\right>$,
$\left<\psi_2\psi_2|\psi_2\psi_2\right>$, and 
$\left<\psi_1\psi_2|\psi_1\psi_2\right>$, in H$_2$ in Figure~\ref{fig:nddoh2_b}. 
We choose to study these $^\psi$ERIs because they are applied in the calculation of the 
$E^\text{(2)}_\text{el}$ ($\left<\psi_1\psi_1|\psi_2\psi_2\right>$) and 
of the full-configuration interaction (FCI)
correlation energy \cite{Szabo2012} (all of these $^\psi$ERIs). 
All of these $^\psi$ERIs are significantly affected by the NDDO approximation.
Hence, it comes as no surprise that the MP2 and FCI correlation energies are 
deteriorated by the NDDO approximation. 
The general shape of $E^\text{(2)}_\text{el}$ follows that of 
$\mathcal{E}_{1122}^{^\psi\text{ERI}}$ and has a minimum at 
$R_{12} = 0.86$ \AA\ where $\mathcal{E}^\text{(2)} = - 0.009$ a.u.
As a consequence, instead of $E^\text{(2)}_\text{el} = -0.016$ a.u. a 
$E^\text{(2)}_\text{el} = -0.007$ a.u.\ is obtained when the NDDO approximation is invoked. 
The MP2 correlation energy is therefore significantly underestimated.
The FCI correlation energy is also significantly underestimated for $R_{12} < 0.72$ \AA. 
An unfortunate addition of the errors in $\left<\psi_2\psi_2|\psi_2\psi_2\right>$  
and $\left<\psi_1\psi_1|\psi_1\psi_1\right>$, however, leads to an 
overestimation of $E^\text{FCI}$ for $R_{12} \ge 0.72$ \AA. 
The errors in the $^\psi$ERIs and, hence, in $E^\text{FCI}$, only vanish in the limit of very large 
$R_{12}$ ($R_{12}> 25.0$ \AA).

\begin{figure}[ht]
 \centering
 \includegraphics[width=\textwidth]{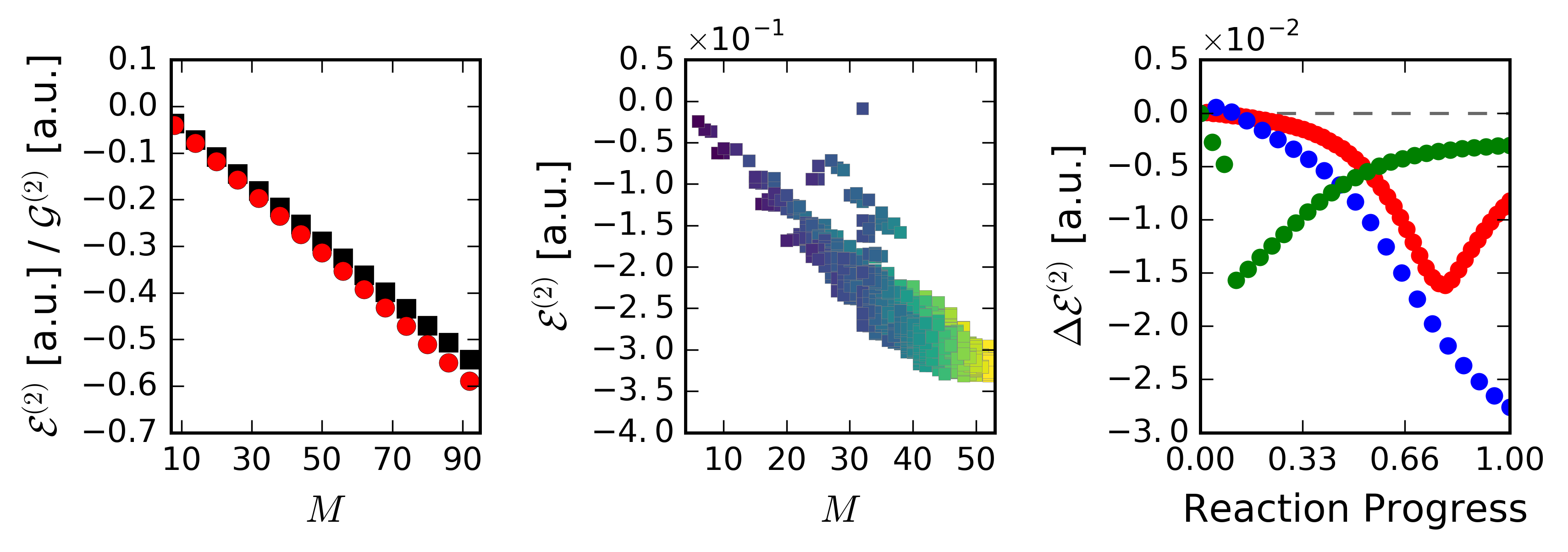}
 \caption{Left: Dependence of $\mathcal{E}^\text{(2)}$ (black squares) and $\mathcal{G}^\text{MP2}$ 
(red circles) on the number of basis functions for molecules in set B. 
Middle:  
Dependence of  $\mathcal{E}^\text{(2)}$ on the number of basis functions for molecules in set C.
The entries for molecule set C are colored according to the 
number of atoms from purple (3 atoms) over blue (12 atoms) 
to yellow (25 atoms).
Right: Dependence of $\mathcal{E}^\text{(2)}$ on the reaction progress  
for reaction A (red circles), reaction E (blue circles), 
and reaction F (green circles). All calculations were carried out with 
the ECP-3G basis set.
}
\label{fig:errorAccumulation3}
\end{figure}

The NDDO approximation also deteriorates the MP2 energies for the other molecules that  
we investigated (see Figure~\ref{fig:errorAccumulation3}). 
Interestingly, the NDDO approximation always produced 
too small MP2 correlation energies. 
The amount by which $E^\text{(2)}_\text{el}$ is underestimated 
depends roughly linearly on the number of basis functions so that $E^\text{(2)}_\text{el}$
is underestimated by $-0.15$ a.u.\ for a molecule with 25 basis functions and 
by $-0.30$ a.u.\ for a molecule with 50 basis functions.
The significance of these results becomes apparent in the context of previous studies
which reported far too small correlation energies determined for NDDO-SEMO reference wave functions. 
\cite{Thiel1981, Clark1993b}
Our results can be taken as an indication 
that the low correlation energies arise as a direct consequence 
of the NDDO approximation and not of the introduction of other parametrized expressions 
when assembling NDDO-SEMO models. 

\section{Analysis of the NDDO Approximation for Molecules in the $\tau$-Basis}
\label{sec:lo}

When applying ordinary $\tau$-basis sets, large errors in the $^\phi$ERIs arise. 
For H$_2$ and $R_{12}=0.74$ \AA\ described in a cc-pVDZ basis set \cite{Dunning1989}, for example, 
the largest absolute error in a $^\phi$ERI amounts to $0.51$ a.u. 
The application of these erroneous $^\phi$ERIs even leads to a $E^\text{HF}_\text{el}=-0.55$ a.u.\ which is 
far too large compared to the exact $E^\text{HF}_\text{el}=-1.13$ a.u.
After the local orthogonalization (Eq.~(\ref{eq:tao2}))
of the basis set, we obtain $\mathcal{G}^\text{HF}=0.01$ a.u.
A prior local orthogonalization led to significantly smaller largest absolute errors in the 
$^\phi$ERIs and cumulative errors in the $^\phi$ERIs for all molecules (B, C, and D) by up to an  
order of magnitude (see also Table~S3 in the Supporting Information). 

\begin{figure}[ht]
 \centering
 \includegraphics[width=.5\textwidth]{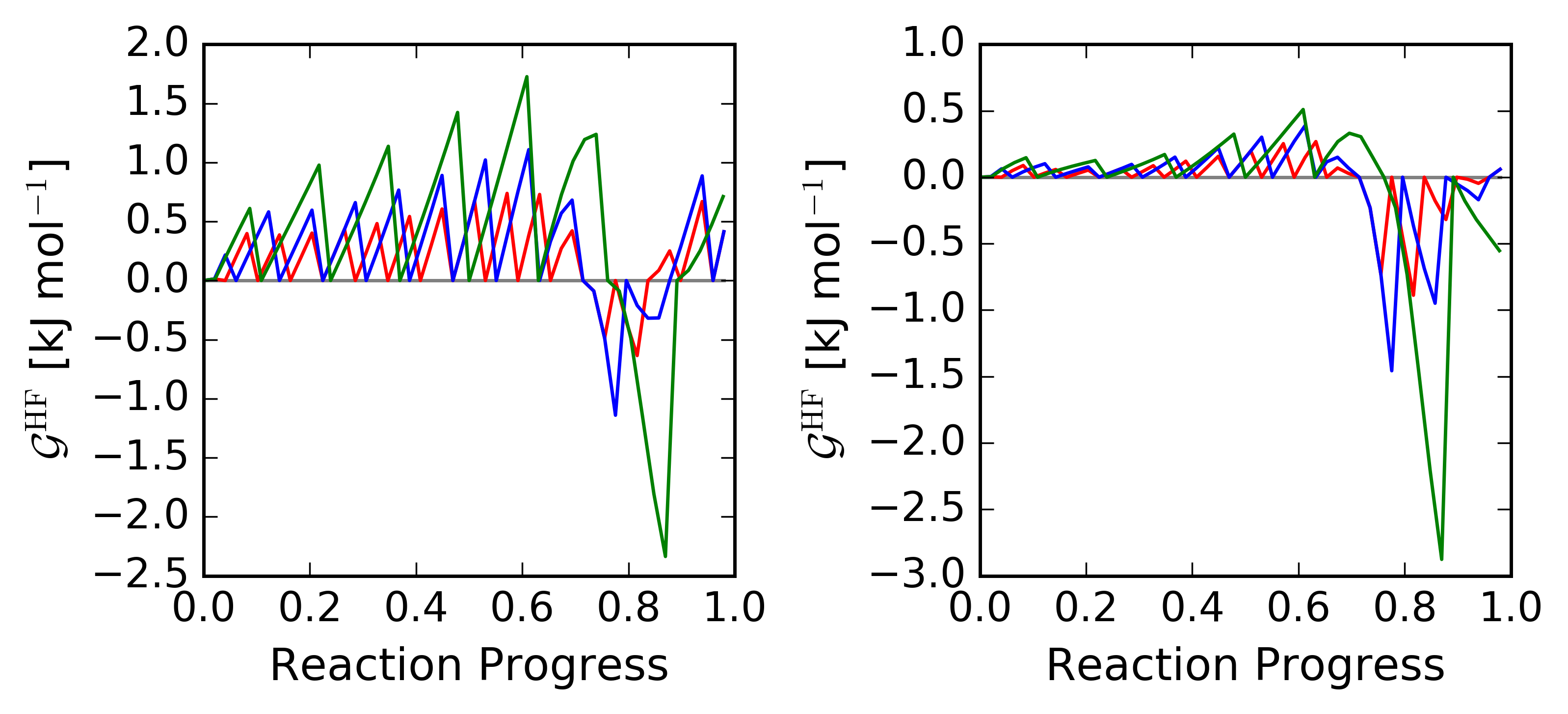}
 \caption{$E^\text{HF}_\text{el}$ for H$_2$ for $R_{12}=0.74$ \AA\  in a.u.\ calculated 
without (black lines) and with (red and blue lines) the NDDO approximation and   
different basis sets. We calculated $E^\text{HF}_\text{el}$ with the NDDO approximation 
either self-consistently (red lines) or non-self-consistently, i.e., with orbitals 
taken from the HF reference calculation (blue lines). 
For reference, we provide the  
HF-limit energy $E^\text{HF}_\text{el}=-1.133629$ a.u. \cite{Jensen1999}
}
\label{fig:basisSetConvergence}
\end{figure}

We found a fundamental limitation of the NDDO approximation 
in the course of our analysis of different $\tau$-basis sets, i.e., 
cc-pVXZ (X = D, \cite{Dunning1989} T, \cite{Dunning1989} Q, \cite{Dunning1989} 5, \cite{Dunning1989} 
6 \cite{emsl}, see Figure~\ref{fig:basisSetConvergence}): 
Usually, $E^\text{HF}_\text{el}$ converges smoothly to the HF limit when larger and larger basis 
sets are applied. The HF limit for H$_2$ for $R_{12}=0.74$ \AA\ was determined to be 
$E^\text{HF}_\text{el}=-1.133629$ a.u. \cite{Jensen1999} 
When applying a cc-pV6Z basis set, we obtain $E^\text{HF}_\text{el}=-1.133476$. 
Figure~\ref{fig:basisSetConvergence} shows that the HF energies calculated with the 
NDDO approximation do not converge with respect to the basis-set size. 
Furthermore, we obtained $E^\text{HF}_\text{el}$ that are smaller than the HF limit which is 
worrisome. 

\section{Improving on the NDDO Approximation}
\label{sec:improve}

The NDDO approximation 
 introduces significant errors in the $^\phi$ERIs, observed here for the simplest possible 
neutral molecule, H$_2$, and for a diverse selection of medium-sized organic compounds.
Obviously, these errors are too large and too unsystematic for the NDDO approximation to  
be useful in purely \textit{ab initio} electronic structure models. 

\subsection{Modifications}
\label{subsec:correcteris}

In \textit{ab initio} electronic structure calculations, it is customary to 
apply screening techniques to determine which $^\chi$ERIs are negligibly small. \cite{Helgaker2014}
Errors are controlled by thresholds with respect to which  
the $^\chi$ERIs are neglected. \cite{Helgaker2014}
By contrast, the NDDO approximation cannot be applied as a screening tool for $^\chi$ERIs
because it can only predict whether the corresponding ERIs in the $\phi$-basis 
are negligibly small or not. 
Each $^\phi$ERI, however, encodes information on \textit{all} $^\chi$ERIs 
(Eq.~(\ref{eq:exactTransformation}). 
It is therefore computationally difficult to improve on the 
NDDO approximation by an explicit transformation of the ERIs from the $\chi$- to 
the $\phi$-basis because this is a 4-index transformation that scales as $\mathcal{O}(M^5)$. 
Some attempts \cite{Kollmar1995,Tu2003}
were made to improve on the approximation of the $^\phi$ERIs for which the corresponding $^\chi$ERIs are 
one-center ERIs 
($\left<\chi_\mu^I \chi_\nu^I | \chi_\lambda^I \chi_\sigma^I\right>$) and two-center ERIs 
($\left<\chi_\mu^I \chi_\nu^I | \chi_\lambda^J \chi_\sigma^J\right>$ for $I\ne J$).
However, these suggestions did not find widespread use. 
Moreover, our results suggest that a correction of these $^\phi$ERIs does not suffice to obtain 
a reliable HF energy $E^\text{HF}_\text{el}$. 
A large portion of the error originates from the $^\phi$ERIs that are assumed to be zero, but 
are not exactly zero. 
For molecule set C, for example, these neglected $^\phi$ERIs on average 
make up 99\% of the $^\phi$ERIs and these are responsible for 81\% of the cumulative absolute error. 
If we had to estimate these $^\phi$ERIs, we would again need close to $\mathcal{O}(M^4)$
operations. These estimates would have to be accurate due to the plethora of 
small $^\phi$ERIs that would again compromise the computational efficiency of the 
NDDO approximation.

\subsection{Capitalizing on Error Cancellation}

In all popular NDDO-SEMO models, the one-center $^\chi$ERIs 
$\left<\chi_\mu^I \chi_\nu^I | \chi_\lambda^I \chi_\sigma^I\right>$ are substituted 
for empirical parameters.
These parameters are usually chosen to be smaller 
than the corresponding analytical one-center $^\chi$ERI. \cite{Dewar1977}
The two-center $^\chi$ERIs 
$\left<\chi_\mu^I \chi_\nu^I | \chi_\lambda^J \chi_\sigma^J\right>$ for $I\ne J$ 
are scaled so that they 
are also smaller than the corresponding analytical $^\chi$ERIs \cite{Dewar1976,Dewar1977} 
(see also Figure~S2 in the Supporting Information). 
In MNDO-type models, the two-center $^\chi$ERIs are evaluated from a classical multipole 
expression truncated after the quadrupole contribution 
which was shown to have a negligible effect on the values of the 
two-center $^\chi$ERIs. \cite{Dewar1976}
Because of the application of parametrized expressions to evaluate the one- and two-center 
$^\chi$ERIs, the results obtained with \textit{ab initio} electronic structure methods invoking 
 the NDDO approximation cannot be directly compared  
to results obtained with modern NDDO-SEMO models.

Furthermore, all successful NDDO-SEMO models introduce various 
parametric expressions to assemble the one-electron matrix 
and to calculate the nucleus-nucleus repulsion energy. \cite{Dewar1977, Stewart1989, Stewart2007,
Stewart2012, Kolb1993a, Weber2000a, Dral2016b}
For contemporary NDDO-SEMO models, the parametric expressions were designed such 
that the result of the SCF optimization yields a result close to a reference energy \textit{despite} 
significant errors in the $^\phi$ERIs compared to the analytical analogues. 
Overall, the results obtained with NDDO-SEMO models achieve a remarkably high accuracy 
with respect to these reference data. \cite{Korth2011, Dral2016a}
At the same time, NDDO-SEMO models are notoriously unreliable for molecules 
not considered in the parametrization procedure. \cite{Korth2011, Dral2016a}
It is virtually impossible to rationalize why errors occur due to the number and the 
diversity of the approximations invoked in an NDDO-SEMO model. 
At least some of these errors are likely to be due to the 
NDDO approximation. 

The parameters in popular NDDO-SEMO models were determined in such a way that the SCF results
deviate in a least-squares manner from experimental reference data. 
In line with the results reported in Section~\ref{subsec:mp2}, the MP2 correlation energy 
$E^\text{(2)}_\text{el}$ obtained with respect 
to a HF-type reference wave function obtained with existing NDDO-SEMO models 
will generally be too small. \cite{Thiel1981, Clark1993b} 
The $^\chi$ERI scaling applied in NDDO-SEMO models 
(which makes the $^\chi$ERIs artificially smaller (see also Figure~S2 in the Supporting Information) 
will worsen the situation because 
the already too small $^\psi$ERIs will become even smaller.
Nevertheless, it might be possible to define 
a NDDO-SEMO model where the error in $E^\text{(0)}_\text{el} + E^\text{(1)}_\text{el}$ 
compensates for 
errors in a subsequent separately calculated $E^\text{MP2}_\text{el}$ in order to justify scaled ERIs
(see also Eq.~(\ref{eq:defmp2})).
A first step in the direction of designing a SEMO model capable of reaching a
satisfactory agreement with coupled cluster electronic energies was 
presented by Margraf \textit{et al.} \cite{Margraf2016}
This model is, however, restricted to single atoms and additional effort would 
be necessary to design a general-purpose NDDO-SEMO model.
The focus of this work, however, is the effect of the NDDO approximation on the ERIs and 
we therefore directly compared ERIs with and without NDDO.

In general, we may anticipate that the improvement 
of the parametric expressions in the NDDO-SEMO models 
is as complicated as the direct correction for the error introduced by the 
NDDO approximation as discussed in Section~\ref{subsec:correcteris}.
Therefore, the only viable use of NDDO-SEMO models appears to be their combination with system-focused 
rigorous error estimation schemes as proposed in Refs.~\onlinecite{Simm2016,Simm2018}.

\subsection{Correcting the Two-Electron Matrices}

We propose an alternative correction strategy, 
the correction inheritance to semiempirics (CISE) approach, which  allows for rapid 
calculations invoking the NDDO approximation with error control.
In 1969, Roby and Sinano\v{g}lu made an attempt to accelerate single-point HF 
calculations for a diverse set of small molecules. \cite{Roby1969}
They suggested to scale $^\chi\fett{G}^{\text{NDDO}}$ 
with a matrix $\fett{\Gamma}$ to obtain a better estimate for $^\phi\fett{G}$,
\begin{equation}
^\phi\fett{G} \approx \fett{\Gamma} {}^\chi\fett{G}^{\text{NDDO}}.
\end{equation}
Their attempt to define general rules to assemble $\fett{\Gamma}$ 
 turned out to be impossible. \cite{Roby1969}

\begin{figure}[ht]
 \centering
 \includegraphics[width=.85\textwidth]{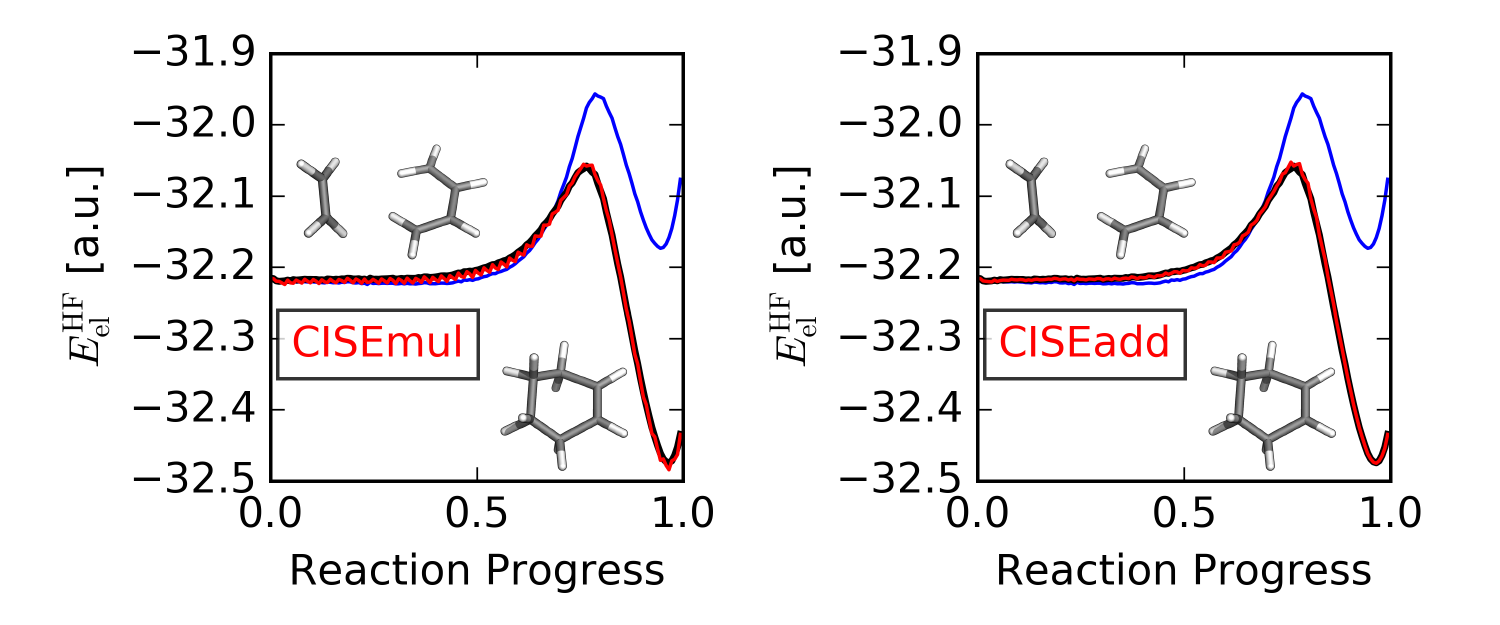}
 \caption{$E^\text{HF}_\text{el}$ for reaction A and the OM2-3G\cite{Stevens1984,Kolb1993a,Weber2000a,Dral2016b}
 basis set (black line). $E^\text{HF-NDDO}_\text{el}$ with orbitals 
taken from the HF reference calculation and shifted by 2.1 a.u.\ (blue line). 
$E^\text{HF-NDDO}_\text{el}$ determined with the CISEmul approach (Eq. (\ref{eq:correct}), red line left) 
and with the CISEadd approach (Eq. (\ref{eq:correct2}), 
red line right). 
The respective scaling matrices $\fett{\Gamma}$, $\fett{\Gamma_J}$, and $\fett{\Gamma_K}$ are
updated every third step. 
}
\label{fig:rscorrection}
\end{figure}

In this Section, we reconsider and build upon the Roby--Sinano\v{g}lu approach. 
We can exactly determine $\fett{\Gamma}(\{\tilde{\fett{R}}_I^n\})$ 
for a given structure $\{\tilde{\fett{R}}_I^n\}$
from a reference HF, KS-DFT, or multi-configurational SCF calculation 
(yielding the exact ${}^\phi\fett{G}^{\{\tilde{\fett{R}}_I^n\}}(^\phi\fett{C})$),
\begin{equation}
 \begin{split}
   \fett{\Gamma}^{\{\tilde{\fett{R}}_I^n\}}(^\phi\fett{C}) &= 
{}^\phi\fett{G}^{\{\tilde{\fett{R}}_I^n\}}(^\phi\fett{C})
 \cdot 
\left( {{}^\chi\fett{G}^{\text{NDDO}}}^{\{\tilde{\fett{R}}_I^n\}}(^\phi\fett{C})\right)^{-1}. \\
 \end{split}
\end{equation}
Figure~\ref{fig:rscorrection} now shows that $\fett{\Gamma}^{\{\tilde{\fett{R}}_I^n\}}(^\phi\fett{C})$ 
is transferable to a certain degree in a sequence of related structures. That is, 
for two similar structures $\{\tilde{\fett{R}}_I^n\}$ and $\{\tilde{\fett{R}}_I^{(n+1)}\}$ we have  
\begin{equation}
\label{eq:correct}
{}^\phi\fett{G}^{\{\tilde{\fett{R}}_I^{(n+1)}\}}(^\phi\fett{C}) \approx 
\fett{\Gamma}^{\{\tilde{\fett{R}}_I^n\}}(^\phi\fett{C}) \cdot  
{{}^\chi\fett{G}^{\text{NDDO}}}^{\{\tilde{\fett{R}}_I^{(n+1)}\}}(^\chi\fett{C}^\text{NDDO}).
\end{equation}
Eq.~(\ref{eq:correct}) defines a system-focused NDDO model, 
the CISEmul model (`mul' for multiplicative), that can be applied in connection 
with any Fock operator. 

The original Roby--Sinano\v{g}lu approach is not the only one conceivable for the
construction of correction matrices. In fact, a multiplicative correction matrix 
changes matrix elements through a combination of elements of the original matrix. An
additive correction appears easier and more straightforward to achieve the goal of
readjusting individual matrix elements.
We may therefore define  
separate additive corrections with the matrices $\fett{\Gamma_J}$ and 
$\fett{\Gamma_K}$ to $^\chi\fett{J}^{\text{NDDO}}$ and 
to $^\chi\fett{K}^{\text{NDDO}}$, 
\begin{equation} 
\label{eq:correct2}
\begin{split}
{}^\phi\fett{G}^{\{\tilde{\fett{R}}_I^{(n+1)}\}}(^\phi\fett{C}) \approx& 
\fett{\Gamma_J}^{\{\tilde{\fett{R}}_I^n\}}(^\phi\fett{C}) +  
{{}^\chi\fett{J}^{\text{NDDO}}}^{\{\tilde{\fett{R}}_I^{(n+1)}\}}(^\chi\fett{C}^\text{NDDO}) \\ &+
\fett{\Gamma_K}^{\{\tilde{\fett{R}}_I^n\}}(^\phi\fett{C}) +  
{{}^\chi\fett{K}^{\text{NDDO}}}^{\{\tilde{\fett{R}}_I^{(n+1)}\}}(^\chi\fett{C}^\text{NDDO}),
\end{split}
\end{equation}
or a unification of $\fett{\Gamma_J}$ and $\fett{\Gamma_K}$ as a total additive correction
(CISEadd approach where `add' stands for additive). 
The correction matrices $\fett{\Gamma_J}$ and $\fett{\Gamma_K}$ may again be obtained 
from a reference HF, KS-DFT, or multi-configurational SCF calculation so that  
\begin{equation} 
 \fett{\Gamma_J}^{\{\tilde{\fett{R}}_I^n\}}(^\phi\fett{C}) = 
{}^\phi\fett{J}^{\{\tilde{\fett{R}}_I^n\}}(^\phi\fett{C})
 - {{}^\chi\fett{J}^{\text{NDDO}}}^{\{\tilde{\fett{R}}_I^n\}}(^\phi\fett{C})
\end{equation}
and 
\begin{equation} 
 \fett{\Gamma_K}^{\{\tilde{\fett{R}}_I^n\}}(^\phi\fett{C}) = 
{}^\phi\fett{K}^{\{\tilde{\fett{R}}_I^n\}}(^\phi\fett{C})
 - {{}^\chi\fett{K}^{\text{NDDO}}}^{\{\tilde{\fett{R}}_I^n\}}(^\phi\fett{C}). 
\end{equation}

We demonstrate the capabilities of the CISEmul and the CISEadd approach on the example of reaction A in 
Figure~\ref{fig:rscorrection}. 
If we do not correct for the error introduced by the NDDO approximation, 
we obtain large errors in the absolute and relative $E^\text{HF}_\text{el}$.
We would for example erroneously predict that cyclohexene is higher in energy 
than butadiene and ethylene (see the blue 
lines in Figure~\ref{fig:rscorrection}).
When we apply the CISEmul approach,  we see that we 
closely follow the exact $E^\text{HF}_\text{el}$. 
For reaction A, the CISEadd approach leads to smaller errors in general and to 
smoother reaction profiles (see Figure~\ref{fig:rscorrection}). 
In this example, we chose to update 
$\fett{\Gamma}^{\{\tilde{\fett{R}}_I^n\}}(^\phi\fett{C})$ every third step.
For this specific example, this leads to energies within chemical accuracy. 
Obviously, the update frequency is crucial when attempting to gain a maximal  
speed-up at a minimal loss of accuracy. 
One could imagine setting up a measure for the necessary update frequency 
for arbitrary reactions by exploiting structural similarity measures\cite{De2016,Huang2016}
as demonstrated in Ref.~\citen{Simm2018}.

Compared to other strategies which apply the NDDO approximation, 
the CISE approach has the advantage that we maintain complete error control on the resulting model
because we can determine $\fett{\Gamma}^{\{\tilde{\fett{R}}_I^{(n+1)}\}}$ 
for a given molecule with nuclear coordinates $\{\tilde{\fett{R}}_I^{(n+1)}\}$ in case of doubt.
In contrast to existing NDDO-SEMO models, we do not have to carry out any statistical calibration of 
parameters. 
Nevertheless, both correction approaches, the CISEmul and the CISEadd approach, 
suffer from limitations. The nuclear coordinates and also the 
density matrices obviously differ in a sequence of structures. Eqs.~(\ref{eq:correct}) and 
(\ref{eq:correct2}) will only yield sensible results when the change of both quantities remains small.
We are currently exploring the possibility to apply this strategy in practice for sequences 
of structures as they occur during structure optimization, 
in Born--Oppenheimer molecular-dynamics trajectories, or in interactive reactivity studies. 

\section{Conclusions}

The NDDO approximation is a central ingredient for many modern SEMO models.
We studied the effect of the NDDO approximation on the ERIs in the symmetrically orthogonalized basis 
for the simplest possible  model system, H$_2$, and for  a diverse set of molecules.
As expected and in agreement with previous results, \cite{Cook1967,Roby1969,Sustmann1969,Gray1970,
Roby1971, Brown1971,Roby1972,Brown1973,Birner1974,Chandler1980,Duke1981, Weinhold1988,Koch2014,
Neymeyr1995a,Neymeyr1995c,Neymeyr1995d,
Neymeyr1995,
Neymeyr1995b,Tu2003} we found that the NDDO approximation leads to 
significant errors for molecules in their equilibrium structure.
The errors do only vanish in the atomization limit where the overlap between different basis functions 
vanishes. 
The errors in the ERIs in the symmetrically orthogonalized basis increase roughly 
linearly with the number of basis functions.
Additionally, we found that the 
errors in the ERIs in the symmetrically orthogonalized basis depend strongly 
on the arrangement of the atomic nuclei. 
These nonnegligible errors introduced by NDDO may only be alleviated by an efficient error cancellation,
which is in operation in SEMO models but not in HF theory.
For HF calculations, error cancellation is unlikely to occur because of the fact that 
the ERIs in the symmetrically orthogonalized basis are contracted with 
 elements of the density matrix. 

We were then able to dissect how the NDDO approximation affects ERIs 
in the \textit{molecular} orbital basis and, hence, correlation energies.
We found that MP2 correlation energies are underestimated and the 
underestimation increases with the number of basis functions.
This finding explains previous reports \cite{Thiel1981,Gleghorn1982} that correlation energies 
obtained with respect 
to an NDDO-SEMO reference wave function are far too small. 

We proposed a local orthogonalization 
that allowed us to transgress the domain of minimal basis sets and to apply ordinary basis sets in 
conjunction with the NDDO approximation. While we observed a drastic reduction in the 
largest absolute errors in the ERIs in the symmetrically orthogonalized basis,
we discovered another limitation of the NDDO approximation.
Electronic energies calculated with the NDDO approximation do not converge with 
respect to the basis set size so that this solution to the small basis-set restriction of NDDO-SEMO 
models does not pay off.

We then continued to propose how one could still capitalize on the efficiency enabled by the NDDO approximation 
without significant loss of accuracy in a system-focused manner for similar structures which 
we called the correction inheritance to semiempirics (CISE) approach.
Specifically, we proposed a strategy to correct for the errors introduced by the 
NDDO approximation in the two-electron matrices which was inspired by 
work of Roby and Sinano{\v{g}}lu. \cite{Roby1969} 
The two-electron matrix obtained within the NDDO approximation is modified with 
a correction matrix obtained from a reference HF, KS-DFT, or multi-configurational SCF calculation. 
These correction matrices are transferable to a certain degree within sequences of related structures.

\section*{Appendix: Computational Methodology}

All calculations in this work were carried out with a modified version of  \textsc{PySCF} (version 1.4) 
\cite{Sun2017,Sun2015}. 
The ERIs in the $\tau$-basis were transformed to the corresponding ERIs in the 
$\chi$-basis or in the $\phi$-basis with the \textsc{ao2mo} integral transformation module of 
\textsc{PySCF}.
We applied the ECP-3G, \cite{Stevens1984,Kolb1993a} STO-3G,\cite{Hehre1969} cc-pVXZ 
(X = D, \cite{Dunning1989} T, \cite{Dunning1989} Q, \cite{Dunning1989} 5, \cite{Dunning1989} 
6 \cite{emsl}), and def2 \cite{Weigend2005} basis sets in calculations. 

Raymond and co-workers assembled a database considered to be representative of 
chemical space. \cite{Ruddigkeit2012}
We randomly chose 5000 entries of the QM9 data set \cite{Ruddigkeit2012,Ramakrishnan2014} (set C) to 
study the error of the NDDO approximation across a large set of molecules.
Additionally, we worked with linearly growing alkane chains (set B) with the stoichiometry 
C$_x$H$_{2x+2},\ (x=1,2,...,15)$. We include the optimized structures as Supplementary Material.
Finally, we selected three reactions (set D) which we had considered\cite{Muhlbach2016}
for interactive reactivity explorations in the framework of 
real-time quantum chemistry\cite{Marti2009,Haag2011,Haag2013,Haag2014,
Haag2014a,Vaucher2016a,Muhlbach2016,Heuer2018}; they can be found in the Supplementary Material in 
Ref.~\citen{Muhlbach2016}.

\section*{Acknowledgements}

This work was supported by the Schweizerischer Nationalfonds.

\providecommand{\refin}[1]{\\ \textbf{Referenced in:} #1}

\end{document}